\documentclass{SciPost}

\usepackage[scaled=1]{XCharter}
\usepackage[bitstream-charter, expert,cal=cmcal]{mathdesign}
\usepackage{upgreek}

\RequirePackage[varqu,varl]{inconsolata}
\DeclareFontFamily{U}{nxlmi}{}
\DeclareFontSubstitution{U}{nxlmi}{m}{it}
\DeclareFontShape{U}{nxlmi}{m}{it}{
    <-6.3>    nxlmi05
    <6.3-8.6> nxlmi07
    <8.6->    nxlmi0
}{}
\DeclareFontShape{U}{nxlmi}{b}{it}{
    <-6.3>    nxlbmi05
    <6.3-8.6> nxlbmi07
    <8.6->    nxlbmi0
}{}
\renewcommand{\partial}{{\text{\scalebox{1.14}{\usefont{U}{nxlmi}{m}{it}\symbol{64}}}\mspace{1mu}}}

\makeatletter
\g@addto@macro\bfseries{\boldmath}
\makeatother

\usepackage[kerning=true]{microtype}

\PassOptionsToPackage{dvipsnames}{xcolor}
\definecolor{maroon}{RGB}{167,74,74}
\definecolor{magreen}{RGB}{59,125,37}
\definecolor{mablue}{RGB}{59,136,195}
\definecolor{mayellow}{RGB}{242,147,24}
\colorlet{mabrown}{maroon!50!magreen}
\colorlet{maorange}{maroon!50!mayellow}
\colorlet{macyan}{magreen!50!mablue}

\colorlet{red}{maroon}
\colorlet{green}{magreen}
\colorlet{blue}{mablue}
\colorlet{yellow}{mayellow}
\colorlet{brown}{mabrown}
\colorlet{orange}{maorange}
\colorlet{cyan}{macyan}

\definecolor{rred}{RGB}{167,33,74}
\definecolor{bblue}{RGB}{29,136,255}
\definecolor{ppurple}{RGB}{113,84,165}
\definecolor{ppink}{RGB}{255,55,219}

\newcommand{\blue}[1]{\textcolor{mablue}{#1}}
\newcommand{\red}[1]{\textcolor{maroon}{#1}}

\newcommand{\gray}[1]{\textcolor{black!40}{#1}}

\RequirePackage{hyperref}
\hypersetup{colorlinks,
    linkcolor=mablue,
    citecolor=magreen,
    urlcolor=maorange,
    anchorcolor=mablue,
    filecolor=mablue,
    menucolor=mablue}

\makeatletter
\renewcommand\section{\@startsection{section}{1}{\z@}%
  {-3.5ex \@plus -1.3ex \@minus -.7ex}%
  {2.3ex \@plus.4ex \@minus .4ex}%
  {\large\scshape\bfseries}}
\renewcommand\subsection{\@startsection{subsection}{2}{\z@}%
    {-2.3ex\@plus -1ex \@minus -.5ex}%
    {1.2ex \@plus .3ex \@minus .3ex}%
    {\normalsize\scshape\bfseries}}
\renewcommand\subsubsection{\@startsection{subsubsection}{3}{\z@}%
    {-2.3ex\@plus -1ex \@minus -.5ex}%
    {1ex \@plus .2ex \@minus .2ex}%
    {\normalsize\scshape\bfseries}}
\renewcommand\paragraph{\@startsection{paragraph}{4}{\z@}%
    {1.75ex \@plus1ex \@minus.2ex}%
    {-1em}%
    {\normalsize\bfseries}}
\renewcommand\subparagraph{\@startsection{subparagraph}{5}{\parindent}%
    {1.75ex \@plus1ex \@minus .2ex}%
    {-1em}%
    {\normalsize\bfseries}}
\makeatother

\newcommand{\beq}{\begin{equation}}
\newcommand{\eeq}{\end{equation}}

\usepackage{twemojis}
\newcommand{\st}{\texttwemoji{monkey face}}
\newcommand{\jf}{\texttwemoji{fish}}

\newcommand{\tcite}[1]{[\textcolor{green}{\sffamily cite}]}

\usepackage{comment} 

\usepackage{import}
\usepackage[inline]{enumitem}
\usepackage{csquotes}

\RequirePackage[thicklines]{cancel}

\newcommand{\nn}{\nonumber \\}

\let\coloneqq\relax
\let\eqqcolon\relax
\makeatletter
\newcommand*{\coloneqq}{\mathrel{\rlap{\raisebox{0.3ex}{$\m@th\cdot$}}\raisebox{-0.3ex}{$\m@th\cdot$}}=}
\newcommand*{\eqqcolon}{\mathrel{=\llap{\raisebox{0.3ex}{$\m@th\cdot$}}\llap{\raisebox{-0.3ex}{$\m@th\cdot$}}}}
\makeatother
\newcommand{\roteqq}{\text{\rotatebox{90}{$=$}}}

\newcommand{\demeqq}{\overset{!}{=}}
\renewcommand{\geq}{\geqslant}
\renewcommand{\leq}{\leqslant}
\usepackage[new]{old-arrows} 

\usepackage{physics}
\newcommand{\parti}{\mathcal{Z}}

\newcommand{\pd}{\partial} 

\newcommand{\inv}[1]{{{#1}^{-1}}} 
\newcommand{\adj}[1]{{#1^{\dagger}}} 
\newcommand{\half}{\tfrac{1}{2}} 
\newcommand{\Half}{\frac{1}{2}}

\newcommand{\rep}[1]{\cX_{#1,\varepsilon}}
\newcommand{\brep}[1]{\pd\rep{#1}}
\newcommand{\sz}[2]{\upzeta_{#1}^{(#2)}}
\newcommand{\suchthat}{\;\middle|\;}

\renewcommand{\vec}{\boldsymbol}

\newcommand{\w}{\mathbin{\scalebox{0.8}{$\wedge$}}} 
\newcommand{\ex}[1]{\mathrm{e}^{#1}} 
\newcommand{\ii}{\mathrm{i}} 
\newcommand{\chk}[1]{#1^{\smash{\scalebox{.9}[1.4]{\rotatebox{90}{\guilsinglleft}}}}} 

\let\C\undefined

\newcommand{\R}{\mathbb{R}}
\newcommand{\C}{\mathbb{C}}

\newcommand{\Z}{\mathbb{Z}}

\renewcommand{\S}{\mathbb{S}}
\newcommand{\T}{\mathbb{T}}

\newcommand{\disc}{\mathbb{D}}

\newcommand{\B}{\mathbb{B}} 
\newcommand{\set}[1]{\qty{#1}}

\newcommand{\tb}{\mathrm{T}} 

\renewcommand{\H}{\mathrm{H}} 

\renewcommand{\t}[1]{{\text{#1}}}
\newcommand{\nspace}[1][1]{\kern-#1em}

\newcommand{\blob}{\bullet}

\renewcommand{\mapsto}{\longmapsto}

\newcommand{\xto}[1]{\overset{#1}{\to}}
\newcommand{\xmapsto}[1]{\overset{#1}{\mapsto}}

\makeatletter
\renewcommand{\xmapsto}[2][]{\ext@arrow 0599{\mapstofill@}{#1}{#2}}
\def\mapstofill@{\arrowfill@{\mapstochar\relbar}\relbar\rightarrow}
\makeatother

\def \cA {\mathcal{A}}
\def \cB {\mathcal{B}}
\def \cC {\mathcal{C}}

\def \cG {\mathcal{G}}
\def \cH {\mathcal{H}}
\def \cI {\mathcal{I}}

\def \cK {\mathcal{K}}

\def \cQ {\mathcal{Q}}
\def \cR {\mathcal{R}}
\def \cS {\mathcal{S}}

\def \cV {\mathcal{V}}

\def \cX {\mathcal{X}}

\def \cZ {\mathcal{Z}}

\def \sN {\mathscr{N}}

\DeclareMathAlphabet{\mathfrak}{U}{jkpmia}{m}{n}

\def \bbE {\mathbb{E}}

\def \bbG {\mathbb{G}}

\def \bbK {\mathbb{K}}

\def \bbO {\mathbb{O}}

\def \bb1 {{\mathbb{1}}}

\def \sfC {\mathsf{C}}

\def \sfI {\mathsf{I}}
\def \sfJ {\mathsf{J}}

\def \sfS {\mathsf{S}}

\def \sfi {\mathsf{i}}
\def \sfj {\mathsf{j}}

\def \sfm {\mathsf{m}}
\def \sfn {\mathsf{n}}

\newcommand{\BF}{\t{BF}}
\newcommand{\edge}{\t{edge}}
\newcommand{\km}{Kac--Moody}

\newcommand{\hW}{\hat{\mathrm{W}}}
\newcommand{\hV}{\hat{\mathrm{V}}}
\newcommand{\aent}{R} 
\newcommand{\coaent}{\co{\aent}} 

\renewcommand{\b}{\mathrm{b}} 

\newcommand{\Ocl}{\Omega_{\text{cl}}}

\newcommand{\cdd}{\dd^{\dagger}}
\newcommand{\lapl}{\triangle}
\newcommand{\tlapl}{\Box}

\newcommand{\pa}{\partial}

\newcommand{\eom}{\t{\textsc{eom}}}

\newcommand{\inp}[2]{#1\mathbin{\lrcorner}#2} 

\usepackage{cancel}

\newcommand{\co}[1]{{#1}^\mathsf{c}}

\newcommand{\fall}{{{}^\forall\!}}

\newcommand{\h}{\mathrm{h}} 
\newcommand{\cpos}{\mathbf{q}} 
\newcommand{\cmom}{\mathbf{p}} 

\usepackage{tikz-cd}

\RequirePackage{mdframed}
\usepackage{empheq}
\newcommand*\obox[1]{%
    \fcolorbox{white}{blue!17}{\hspace{1em}#1\hspace{1em}}}

\newcommand{\ent}{\cS}
\newcommand{\ee}{\cS_\t{EE}}
\newcommand{\tee}{\cS_\t{TEE}}

\newcommand{\kket}[1]{\left.\ket{#1}\!\right\rangle}
\newcommand{\bbra}[1]{\left\langle\!\bra{#1}\right.}

\def\fdiffd{\mathrm{D}}
\DeclareDocumentCommand\fdifferential{ o g d() }{ 
    \IfNoValueTF{#2}{
        \IfNoValueTF{#3}
        {\fdiffd\IfNoValueTF{#1}{}{^{#1}}}
        {\mathinner{\fdiffd\IfNoValueTF{#1}{}{^{#1}}\argopen(#3\argclose)}}
    }
    {\mathinner{\fdiffd\IfNoValueTF{#1}{}{^{#1}}#2} \IfNoValueTF{#3}{}{(#3)}}
}
\DeclareDocumentCommand\DD{}{\fdifferential}

\DeclareDocumentCommand\variation{ o g d() }{ 
    \IfNoValueTF{#2}{
        \IfNoValueTF{#3}
        {\updelta \IfNoValueTF{#1}{}{^{#1}}}
        {\mathinner{\updelta \IfNoValueTF{#1}{}{^{#1}}\argopen(#3\argclose)}}
    }
    {\mathinner{\updelta \IfNoValueTF{#1}{}{^{#1}}#2} \IfNoValueTF{#3}{}{(#3)}}
}
\DeclareDocumentCommand\var{}{\variation} 

\def\sdiffd{\mathbf{d}}
\DeclareDocumentCommand\sdifferential{ o g d() }{ 
    \IfNoValueTF{#2}{
        \IfNoValueTF{#3}
        {\sdiffd\IfNoValueTF{#1}{}{^{#1}}}
        {\mathinner{\sdiffd\IfNoValueTF{#1}{}{^{#1}}\argopen(#3\argclose)}}
    }
    {\mathinner{\sdiffd\IfNoValueTF{#1}{}{^{#1}}#2} \IfNoValueTF{#3}{}{(#3)}}
}
\DeclareDocumentCommand\sd{}{\sdifferential}

\DeclareDocumentCommand\detp{}{\opbraces{\determinant{}'}}

\DeclareMathOperator{\volume}{vol}
\DeclareDocumentCommand\vol{}{\opbraces{\volume}}

\DeclareMathOperator{\kernel}{ker}
\DeclareDocumentCommand\ker{}{\opbraces{\kernel}}

\DeclareMathOperator{\image}{im}
\DeclareDocumentCommand\im{}{\opbraces{\image}}

\DeclareMathOperator{\spectrum}{spec}
\DeclareDocumentCommand\spec{}{\opbraces{\spectrum}}

\DeclareMathOperator{\character}{ch}
\DeclareDocumentCommand\ch{}{\opbraces{\character}}

\DeclareMathOperator{\harm}{Harm}

\DeclareMathOperator{\homomorphism}{Hom}
\DeclareDocumentCommand\hom{}{\opbraces{\homomorphism}}

\renewcommand{\ip}[2]{\left\langle#1,#2\right\rangle}

\RequirePackage{interval}
\intervalconfig{soft open fences}

\usepackage[noabbrev]{cleveref}

\crefname{subsection}{subsection}{subsections}
\crefname{equation}{}{}
\crefname{enumi}{}{}

\numberwithin{equation}{section}

\usepackage[style=numeric-comp, eprint=true, date=year, natbib, backend=biber, maxbibnames=10, giveninits=true, sorting=none, useprefix=true]{biblatex}
\DeclareFieldFormat*{title}{\emph{#1}}
\DeclareFieldFormat{pages}{#1}
\DeclareFieldFormat{labelalpha}{\textsc{#1}}

\renewbibmacro{in:}{}
\begin{document}

\begin{center}
	{\Large\bfseries\scshape{Entanglement in BF theory II: \\  edge-modes}}
\end{center}

\begin{center}
	\textbf{
		Jackson R. Fliss\textsuperscript{1,\jf} and
		Stathis Vitouladitis\textsuperscript{2,\st}
	}
\end{center}

\begin{center}
	\textbf{1}  Department of Applied Mathematics and Theoretical Physics, \\ University of
	Cambridge, Cambridge CB3 0WA, United Kingdom
	\\[0.5em]
	\textbf{2}  Institute for Theoretical Physics, University of Amsterdam, \\ 1090 GL
	Amsterdam, The Netherlands
	\\[\baselineskip]

	\jf\ \href{mailto:jf768@cam.ac.uk}{\small \sf jf768@cam.ac.uk} \qquad \st\ \href{mailto:e.vitouladitis@uva.nl}{\small \sf e.vitouladitis@uva.nl}
\end{center}

\vspace{4em}
\section*{Abstract}
We consider the entanglement entropy arising from edge-modes in Abelian $p$-form topological field theories in $d$ dimensions on arbitrary spatial topology and across arbitrary entangling surfaces. We find a series of descending area laws plus universal corrections proportional to the Betti numbers of the entangling surface, which can be taken as a higher-dimensional version of the \textquote{topological entanglement entropy.} Our calculation comes in two flavors: firstly, through an induced edge-mode theory appearing on the regulated entangling surface in a replica path integral and secondly through a more rigorous definition of the entanglement entropy through an extended Hilbert space. Along the way we establish several key results that are of their own merit. We explain how the edge-mode theory is a novel combination of $(p-1)$-form and $(d-p-2)$-form Maxwell theories linked by a chirality condition, in what we coin a \emph{chiral mixed Maxwell theory}.  We explicitly evaluate the thermal partition function of this theory. Additionally we show that the extended Hilbert space is completely organized into representations of an infinite-dimensional, centrally extended current algebra which naturally generalizes 2d Kac-Moody algebras to arbitrary dimension and topology. We construct the Verma modules and the representation characters of this algebra. Lastly, we connect the two approaches, showing that the thermal partition function of the chiral mixed Maxwell theory is precisely an extended representation character of our current algebra, establishing an exact correspondence of the edge-mode theory and the entanglement spectrum.

\pagebreak
\tableofcontents
\vspace{10pt}

\section{Introduction}\label{sec:intro}

Long-range entanglement is a notion with potent conceptual and practical utility in characterizing quantum phases of matter. In gapped ground states, short-range correlations manifest in an `area law' entanglement coming from degrees of freedom localized and straddling the boundary of the region of the interest (within a correlation length), the so-called \textquote{entangling surface.} Important, however, are potential long-range corrections to this area law. This is exemplified in (2+1)-dimensional gapped systems where a constant negative correction to the area law can arise from non-local features of the ground state, which constrain the short-ranged correlations at the entangling surface.  This is the celebrated \textquote{topological entanglement entropy} and is a smoking gun of (2+1)d topological order \cite{kitaev2006topological,Levin:2006zz}.

This story is mirrored beautifully in topological quantum field theory (TQFT), which provides IR effective field theories of topological order: when restricted to a region of spacetime, TQFTs are host to a robust spectrum of \textquote{edge-modes} localized to the boundary of the region \cite{Li:2008kda, chandran2011bulk, Swingle:2011hu, qi2012general}. These edge-modes are the inheritance of bulk gauge transformations, which are broken by the existence of a boundary. When \textquote{gluing} a region to its complement to form a complete state, the edge-modes on either side of the common boundary are maximally entangled up to global constraints \cite{Wen:2016snr,Fliss:2020cos}, giving a divergent area-law entanglement entropy with universal constant corrections. TQFT also provides powerful avenues to corroborate these results, such as through the replica trick and surgery \cite{Dong:2008ft,Fliss:2020cos}.

By this point, the role of edge-modes in topological entanglement entropy is extremely well understood in (2+1) dimensions; while generally believed to extend, their role is still relatively unexplored in general dimensions. In this work we make progress in this direction, focussing on the Abelian topological phases described by \(p\)-form BF theories in \(d\) dimensions:
\begin{align}
	S_\BF \coloneqq \frac{\bbK^{\sfI\sfJ}}{2\pi} \int B_\sfI \w\dd A_\sfJ.
\end{align}
Above $A_\sfJ$ is a vector of \(p\)-forms and $B_\sfI$ is a vector of $(d-p-1)$-forms and $\bbK$ is a rank $\kappa$ symmetric matrix\footnote{In principle, this action doesn't require $\bbK$ to be symmetric, however only the symmetric part will participate in the edge mode entanglement and this saves us introducing extra notation.} of integer entries. These are theories whose ground states are \(p\)-form membrane condensates. In a companion paper, we explained how such theories display an extreme long-range form of entanglement, what we name \textquote{essential topological entanglement} (ETE), which must be present in the strict IR limit of a TQFT \cite{Fliss:2023dze}. The ETE is entirely finite: absent from it are all contributions from UV degrees of freedom and their subsequent area law. However, it is still important to know the UV contributions to entanglement entropy which provide benchmarks for simulating topological order with a given UV model\footnote{A real-world material being a pertinent example!} (such as through tensor networks or matrix product states). In this direction, the potential contributions to topological entanglement entropy in \(p\)-form condensates were deduced by Grover, Turner, and Vishwanath (GTV) \cite{Grover:2011fa} by focussing on lattice gauge theories as a particular UV realization. Despite various example calculations,\footnote{See e.g. \cite{Ma:2017igk,Delcamp:2019fdp,Wen:2017xwk,Bridgeman:2020sqz,Bullivant:2015tva,Zheng:2017yta}. See \cite{Ibieta-Jimenez:2019wwo} for a very general lattice gauge theory calculation.} to date there has not been a comprehensive computation performed in the corresponding continuum TQFT for generic entangling surfaces and any dimension. In this paper, we fill this gap.

We will explain how the edge-mode spectrum leads to area (plus subleading area) law entanglement entropy with constant pieces that are sensitive to topological features of the entangling surface. To be concrete, we find
\begin{align}\label{eq:mainresult1}
	\ee & = \sum_{k=1}^{\left\lfloor\frac{d-1}{2}\right\rfloor} \mathrm{C}^{(p-1)}_k \qty(\frac{\ell}{\varepsilon})^{d-2k} + \frac{\kappa}{2} \qty(\cI_{\frac{d-2}{2}}^{(p-1)}+\cI_{\frac{d-2}{2}}^{(d-p-2)}) \delta_{d,\t{even}}\log(\frac{\ell}{\varepsilon}) \nonumber \\
	    & \phantom{=~} - \frac{1}{2}\qty(\b_p+\b_{d-p-1})\log\abs{\det\bbK}.
\end{align}
where $\varepsilon$ is a short-distance regulator, $\mathrm C^{(p-1)}_k$ are non-universal dimensionless numbers, that we compute, \(\cI_{\frac{d-2}{2}}^{(k)}\) is the \(\qty(\frac{d-2}{2})\)-th heat kernel coefficient for a particular spectral zeta function, $\ell$ is a characteristic length scale, and $\b_n$ denotes the $n$-th Betti number of the entangling surface.  This is the first main result of this paper. Our result differs from the GTV result; in the discussion, \Cref{sec:disc}, we discuss to what extent the two results can be reconciled through local terms to the entangling surface. We arrive at this result through two independent, yet conceptually complementary, avenues.
\begin{enumerate}[label=(\Roman*)]
	\item\label{item:replica} Firstly we utilize the replica trick and perform path integral on a replica manifold, regulated by excising a small region (of circumference $\varepsilon$) about the entangling surface. The resulting edge-mode theory on this regulated entangling surface is a novel mix of $(p-1)$-form and $(d-p-2)$-form Maxwell theories, tied together by a \textquote{chirality} condition. We coin this theory \textquote{chiral mixed Maxwell theory.} The existence of such theories as well as their thermal partition function, which we explicitly compute, is a second main result of this paper. This edge-mode theory contributes an entropy given by \Cref{eq:mainresult1}.

	\item\label{item:KM} Secondly, and more rigorously, we address the subtle issues of gauge-invariance in defining entanglement entropy by moving to an \textquote{extended Hilbert space} (EHS) \cite{Buividovich:2008gq,Donnelly:2011hn,Donnelly:2014fua,Donnelly:2015hxa,Soni:2015yga}. We show that the EHS is organized by an infinite-dimensional current algebra, akin to the \km\ algebras that arise at the edge of (2+1) dimensional phases. This algebra has also appeared in the context of 4d Abelian Maxwell theory \cite{Hofman:2018lfz}, where fixes the spectrum of the theory and can, for instance, be used to establish a state-operator correspondence for nonlocal operators \cite{Hofman:2024oze}.  However, to our knowledge, the existence of these algebras for general \(p\)-forms and in general dimensions has not been explored. Here we will elucidate them in detail, construct their Verma modules, and compute their representation characters. We regard this as a third main result of this paper.  Unlike in (2+1)d, these algebras are not necessarily conformal. Regardless, they completely fix the computation of \Cref{eq:mainresult1} which arises from the high-temperature limit of the representation character of this algebra (i.e., a regulated count of the representation dimension).
\end{enumerate}

Along the way, we explain the connection between these two approaches: the chiral mixed Maxwell theory appearing in \Cref{item:replica} has a spectrum that is completely fixed by the infinite-dimensional algebras of approach \Cref{item:KM}. Correspondingly, their partition function is given exactly by a representation character, which is the ultimate source of the match in \Cref{eq:mainresult1}. This is a precise analogue of \textquote{edge spectrum = bulk entanglement spectrum} promoted to higher dimensions. We regard this as a final major result of this paper.

An organizing summary of the paper is as follows. In \Cref{sec:BF-PI} we introduce the BF theory and perform its path integral on manifolds with and without boundary.  While there are many results for the BF path integral on closed manifolds, we will be very careful keeping track of factors of $\bbK$, which, to our knowledge, had not been fully nailed down in the previous literature.  We will use these results in \Cref{sec:replica} to evaluate the replica integral and describe how the computation ultimately results from the partition function of the chiral mixed Maxwell theory. In \Cref{sect:edgemodes} we shift gears and describe the more systematic definition of the entanglement entropy through the extended Hilbert space and describe features of the resulting current algebra that organizes it. We then use the representation characters of this algebra to compute the entanglement entropy, finding a match with \Cref{sec:replica}. In the discussion, \Cref{sec:disc}, we will put our results in context with the known results of GTV, as well as the ETE of this theory, and possible future extensions of our computation.

\subsection{Notation}

For the reader's ease, we introduce here some basic notation that will be used in what follows.

We will analyze theories on torsion-free manifolds of spacetime dimension \(d\), which we shall collectively denote as \(X\). These manifolds may have a boundary, which we will embed into \(X\) using the map $i_\pd:\pd X\hookrightarrow X$. Theories will be quantized on manifolds of dimension $D\equiv d-1$, which we will refer to as $\Sigma$, often calling it the \textquote{Cauchy slice} without reference to any causal structure of the TQFT. We will consider a subregion $\aent$, which is the closure of a \(D\)-dimensional embedded open submanifold of $\Sigma$. The interior of $\aent$ is $\overline{\aent}\coloneqq\aent\setminus\pa\aent$, and its complement is $\coaent$, the closure of $\Sigma\setminus\aent$. Note that $\aent\cap\coaent=\pa\aent$. We will denote the space of forms of degree \(p\) as $\Omega^p(\;\cdot\;)$. Unless stated otherwise, these forms will be real valued. Cohomology groups will be denoted with their degree placed upstairs, $\H^p(\;\cdot\;)$, while homology groups will be denoted with their degree placed downstairs, $\H_p(\;\cdot\;)$, and these groups are always defined with integer coefficients, unless stated otherwise. For compact, boundary-less manifolds, we notate the dimensions of the groups by the Betti number, i.e.:
\begin{equation}
	\b_p(\;\cdot\;)\coloneqq\dim\H^p(\;\cdot\;)=\dim\H_p(\;\cdot\;).
\end{equation}
For (co)homology groups on manifolds with boundary or for relative homology groups, we will always explicitly write the dimension.

\section{The BF path integral}\label{sec:BF-PI}

In this section, we describe the path integral quantization of multi-component, Abelian, \(p\)-form BF theory, on a \(d\)-dimensional, torsion-free manifold, \(X\), with a potentially nonempty boundary, $\pd X$. We will start by reminding the reader of the procedure in the case where \(X\) is a closed manifold, presenting general results with the added benefit of a careful accounting of the level matrix. We then move on to modify the situation in the presence of boundaries.

\subsection{On a closed manifold}\label{ssec:closed}

Let us set the stage by starting with a closed manifold. We consider BF theory defined by the action
\begin{align}\label{eq:BFact}
	S_\BF[A,B] \coloneqq \frac{\bbK^{\sfI\sfJ}}{2\pi} \int_X B_\sfI \w\dd A_\sfJ.
\end{align}
In the above, \(A_\sfI\in \Omega^p(X)\) and \(B_\sfJ\in \Omega^{d-p-1}(X)\) are vectors of \(p\)- and \((d-p-1)\)-form gauge fields respectively. The level matrix, \(\bbK\) is a symmetric, integer, and non-degenerate matrix of rank \(\kappa\). In what follows, we will drop the indices, wherever not necessary, to simplify the notation. Note that, as emphasized above, this definition makes sense whenever \(X\) is torsion-free. We will comment on the case of torsion manifolds at the end of this subsection. For a more general and precise definition of BF theory, we refer the reader to \cite[Appendix A]{Fliss:2023dze}.

The equations of motion arising from varying the action \Cref{eq:BFact} are flatness conditions:
\begin{align}
	\eom[A]=\frac{\bbK}{2\pi}\dd{A}=0 \qquad\qquad \eom[B]=(-1)^{(d-p)(p+1)}\frac{\bbK}{2\pi}\dd{B}=0.
\end{align}
Moreover, the action \Cref{eq:BFact} possesses a gauge redundancy of the form
\begin{equation}
	\var A = \alpha \qq{and} \var B = \beta,
\end{equation}
where \(\alpha\in \Ocl^p(X)\) and \(\beta\in \Ocl^{d-p-1}(X)\), are closed \(p\)- and \((d-p-1)\)-forms respectively. Note that there are two types of gauge shifts. Shifts by harmonic forms correspond to large gauge transformations, in the sense that they are not continuously connected to the identity, and shifts by exact forms correspond to the usual infinitesimal gauge transformations. To properly quantize the theory in the path integral formalism, we must therefore divide by the volume of the gauge groups. However, note that there is a tower of reducibility of the gauge parameters. For example, splitting \(\alpha\) off as \(\alpha = \alpha_0+\dd\alpha_1\), with \(\alpha_0\in\harm^p(X)\) and \(\alpha_1\in \Omega^{p-1}(X)\), the parameter \(\alpha_1\) generates the same gauge transformation as \(\alpha_1+\tilde{\alpha}_1\), with \(\tilde{\alpha}_1\in \Omega^{p-1}_\t{cl}(X)\). This tower continues until one reaches a zero-form gauge redundancy. This redundancy is encoded in the BF path integral in the form of the volume of the gauge group:
\begin{align}
	\parti_\BF[X] = \int \frac{\DD{A}\,\DD{B}}{\vol(\cG)}\ \ex{\ii S_\BF[A,B]},
\end{align}
where, $\cG$ is the total gauge group, $\cG=\cG_p\times\cG_{d-p-1}$, with
\begin{align}\label{eq:gauge-group}
	\cG_k \coloneqq \Omega^{k}_\t{cl}(X)\big/\cG_{k-1},
\end{align}
and \(\cG_0 = \harm^0(X)\). In what follows we will be careless with overall numerical coefficients, such as factors of $2\pi$ in the partition function, as they can be absorbed in an overall normalization of the path integral measure. However, we will pay extra attention to factors of $\bbK$, as they play a crucial role in the universal terms of the entanglement entropy.

One commonly employed method to properly quantize such higher-gauge theories is to include ghosts and ghosts-for-ghosts, and so on, until the gauge transformations are completely resolved \cite{Blau:1989bq}. An alternative approach, due to \cite{Gegenberg:1993gd}, which was shown to be equivalent, at the level of determinants, to that of Blau and Thompson  \cite{Blau:1989bq} (which is in turn also equivalent to Schwarz's method of resolvents \cite{Schwarz:1979ae}), is to Hodge-decompose the fields as
\begin{equation}\label{eq:BA-hodge-decomp}
	\begin{split}
		A &= A_0 + \dd{A_\perp} + \cdd{A_\parallel}, \\
		B &= B_0 + \dd{B_\perp} + \cdd{B_\parallel}.
	\end{split}
\end{equation}
with \(A_0\in\harm^p(X)\), \(A_\perp\in \Omega^{p-1}(X)\), \(A_\parallel\in \Omega^{p+1}(X)\), \(B_0\in\harm^{d-p-1}(X)\), \(B_\perp\in \Omega^{d-p-2}(X)\), and \(B_\parallel\in \Omega^{d-p}(X)\) and perform the path integral directly. In what follows, we will show that, with a bit of care, this method also correctly reproduces the harmonic correction to the partition function\footnote{This corresponds to the refinement of the Ray--Singer torsion as an element of the determinant line bundle, \(\det\H^\blob\),  of \(\H^\blob(X)\).} and we will obtain the level dependence of the partition function. With the decomposition \Cref{eq:BA-hodge-decomp}, taking into account the Jacobians of the transformation, the path integral measure decomposes as
\begin{align}
	\DD{A} = \DD{A_0}\ \DD{A_\perp} \DD{A_\parallel}\ \qty(\detp_{\Omega^{p-1}(X)}(\cdd\dd))^{1/2}\qty(\detp_{\Omega^{p+1}(X)}(\dd\cdd))^{1/2},
\end{align}
and similarly for \(\DD{B}\), while the action takes the form
\begin{align}
	S_\BF[A,B] = \frac{\bbK}{2\pi}\ip{\cdd B_\parallel}{\star\dd\cdd A_\parallel},
\end{align}
where \(\ip{\red\blob}{\blue\blob}\) is the Hodge inner product, \(\int_X \red\blob\w\star\blue\blob\). Following \cite{Donnelly:2016mlc} we will normalize the path integral measure as
\begin{align}\label{eq:measure}
	\DD{A} = \prod_{x\in X} \bbE \dd{A_x} \qquad\qquad \DD{B} = \prod_{x\in X} \bbE \dd{B_x},
\end{align}
where \(\bbE\) is a frame matrix, satisfying \(\bbE^2 = \bbK\). This has the effect of removing the \(\bbK\) dependence from the functional determinants, arising upon integrating over \(A_\parallel\) and \(B_\parallel\). Note that this is not the only choice. Normalizing the modes as $\sim \bbK^\alpha \dd{A_x}$ and $\sim \bbK^{1-\alpha}\dd{B_x}$, for some real $\alpha$, still takes care of the $\bbK$-dependence of the functional determinants, but as will become evident below, it rescales the partition function as
\begin{equation}
	\parti_\BF[X]\mapsto\abs{\det\bbK}^{\alpha\,\upchi(X)}\parti_\BF[X],
\end{equation}
where $\upchi(X)$ is the Euler characteristic of \(X\). This ambiguity is on the one hand, well documented and understood \cite{Witten:1991we} and on the other hand harmless if one is interested in topological entanglement entropy. The reason is, ultimately, that it can be seen to arise from (or equivalently can be absorbed into) a locally integrated contribution \cite{Grover:2011fa}. We will return to this point in the discussion, \Cref{sec:disc}.

Ignoring this ambiguity, for the reasons mentioned above, the integral over $A_\parallel$ and $B_\parallel$ with the measure \Cref{eq:measure} produces
\begin{equation}\label{eq:Zparallel}
	\parti_\BF^\parallel[X] = \qty(\detp_{\Omega^{p}(X)}(\cdd\dd))^{-\frac{3}{2}}.
\end{equation}
Integrating over \(A_\perp\) and \(B_\perp\), modulo small gauge transformations, i.e. the part of \Cref{eq:gauge-group} generated by \(\dd\Omega^{k-1}(X)\), gives an alternating product of determinants, which can be combined with \Cref{eq:Zparallel} to give the analytic torsion\footnote{Actually we get this expression for odd-dimensional manifolds. For even-dimensional manifolds we get a different power of the (determinant expression of the) analytic torsion. However, on even-dimensional manifolds the analytic torsion is unity, so we can write \Cref{eq:Znonzero} for all dimensions.} \cite{Gegenberg:1993gd}:
\begin{align}\label{eq:Znonzero}
	\parti_\BF^\perp[X]\parti_\BF^\parallel[X] = \prod_{k=0}^d \qty(\detp_{\Omega^k(X)}\lapl_n)^{\frac{k}{2}(-1)^{p+k}} \eqqcolon \tb_\t{A}[X]^{(-1)^{p-1}}.
\end{align}
As it stands, \(\tb_\t{A}[X]\) depends on the metric, whenever not all cohomology groups are trivial. So this cannot be the final expression of the partition function; it is the zero-modes that will provide the fix, as is usually the case. Let us see what their contribution is. The integral over \(A_0\) and \(B_0\), modulo large gauge transformations, with the measure \Cref{eq:measure} gives \cite{Donnelly:2016mlc,blauMassiveRaySingerTorsion2022}:
\begin{align}
	\parti_\BF^0[X] & = \qty(\prod_{k=0}^p \abs{\det\qty(\bbK\otimes \bbG_k)}^{\Half(-1)^{p-k}})\qty(\prod_{\ell=0}^{d-p-1} \abs{\det\qty(\bbK\otimes \bbG_\ell)}^{\Half(-1)^{d-p-1-\ell}}),
\end{align}
where \(\bbG_k\) is the metric on the moduli space of harmonic \(k\)-forms, i.e. the Gram matrix of the topological basis of harmonic \(k\)-forms. More explicitly, the topological basis, \(\set{\tau^{(k)}_\sfi}_{\sfi=1}^{\b_k(X)}\), is defined by the relation
\begin{align}\label{eq:top-basis-def}
	\int_{C^\sfi_{(k)}} \tau^{(k)}_\sfj = \delta^\sfi_\sfj,
\end{align}
for a fixed basis, \(\set{C^\sfj_{(k)}}_{\sfi=1}^{\b_k(X)}\), of \(k\)-cycles. Then \(\bbG_k\) is defined as
\begin{align}\label{eq:Gram}
	\mqty[\bbG_k]_{\sfi\sfj} \coloneqq \ip{\tau^{(k)}_\sfi}{\tau^{(k)}_\sfj}.
\end{align}
Poincaré duality implies that (up to a unimodular matrix, which can be set to one by a suitable choice of basis of the \(k\)-cycles) \(\bbG_{d-k}=\inv{\bbG_k}\). Utilizing this, we get:
\begin{align}\label{eq:Z0}
	\parti_\BF^0[X] = \abs{\det\bbK}^{\h_{p}(X)}  \tb_{\H^\blob}[X]^{(-1)^{p-1}},
\end{align}
where
\begin{align}\label{eq:hp}
	\h_{p}(X) & \coloneqq \Half(-1)^p\sum_{k=0}^p(-1)^k\b_k(X) + \Half(-1)^{d-p-1}\sum_{k=0}^{d-p-1}(-1)^k\b_k(X) \nn
	          & \overset{\upchi}{=} (-1)^p\sum_{k=0}^p(-1)^k\b_k(X),
\end{align}
where the last equality is modulo removing factors of the Euler characteristic, which can be done by adding a local counterterm, as we alluded to above. Moreover, we have defined
\begin{align}
	\tb_{\H^\blob}[X] \coloneqq \prod_{k=0}^d \qty(\det\bbG_k)^{\Half(-1)^{k+1}}.
\end{align}
\(\tb_{\H^\blob}[X]\) is precisely the cohomological correction to the analytic torsion, necessary to form the metric-independent Ray--Singer torsion:
\begin{align}
	\tb_\t{RS}[X] = \tb_{\H^\blob}[X] \tb_\t{A}[X].
\end{align}

Putting everything together, the partition function for multi-component, \(p\)-form BF theory on a compact, closed manifold \(X\), reads:
\begin{empheq}[box=\obox]{equation}\label{eq:ZBF}
	\parti_\BF[X] = \abs{\det\bbK}^{\h_{p}(X)}\; \tb_\t{RS}[X]^{(-1)^{p-1}}.
\end{empheq}
This form of the partition function has a number of desirable features which we list below.
\begin{itemize}
	\item It is topological. Of course, this was anticipated since the very first appearance of the theory, in \cite{Schwarz:1979ae}. However, to obtain it in the path integral formalism requires some care, regardless of the method one chooses. The method of resolutions \cite{Schwarz:1979ae} was unable to reproduce it, in cases with non-trivial cohomology groups. In the BRST formulation, one needs to append the action by a BRST-closed, but not necessarily BRST-exact, quantum action \cite{Blau:1989bq,blauMassiveRaySingerTorsion2022}. In the Batalin--Vilkovisky (BV) \cite{Batalin:1983ggl} formulation of the theory \cite{Cattaneo:2015vsa} one needs to pay special attention to the residual superfields, furnishing the fiber of the BV integral. Finally, in the direct method that we employed one must be carfeul about the zero-modes appearing in the tower of gauge-for-gauge-for-\ldots-gauge volumes; note, for example, that in \cite{Gegenberg:1993gd} the zero-mode piece is not correctly reproduced and hence the formulas there depend implicitly on a choice of metric on \(X\).
	\item On a three-dimensional manifold, with \(p=1\), \Cref{eq:ZBF} is the square of the partition function of Abelian Chern--Simons theory. More explicitly, on \(\S^3\) we obtain:
	      \begin{align}
		      \parti_\BF\qty[\S^3] = \abs{\det\bbK}^{-1},
	      \end{align}
	      which is to be compared with \(\parti_\t{CS}[\S^3] = \abs{\det\bbK}^{-1/2}\) in \cite{Fliss:2017wop}. For a more general case, in \(d=3\) with \(p=1\), allowing also for torsion, see also \cite{Mathieu:2015mda}.
	\item On \(X=\S^1\times \Sigma\), \(\parti_\BF[\S^1\times\Sigma]\) simply counts the number of ground states on \(\Sigma\). Equivalently, it measures the dimension of the Hilbert space on \(\Sigma\). This is given by \linebreak \(\dim\cH_\Sigma = \abs{\det\bbK}^{\b_p(\Sigma)}\) \cite{Fliss:2023dze}. From \Cref{eq:ZBF}, making use of the Künneth formula: \linebreak \(\b_k(X) = \b_k(\Sigma)+\b_{k-1}(\Sigma)\), if \(k\geq 1\), while \(\b_0(X)=\b_0(\Sigma)\), we correctly find
	      \begin{equation}\label{eq:BFHSdim}
		      \parti_\BF\qty[\S^1\times \Sigma] = \abs{\det\bbK}^{\b_p(\Sigma)}.
	      \end{equation}
\end{itemize}

Lastly, let us mention that if we allow \(X\) to have torsion, in which case we should take into account the differential cohomology definition of BF theory \cite[Appendix A]{Fliss:2023dze}, the expression for the partition function is modified as follows. Let \(\H_p(X;\Z) = \Z^{\b_p(X)}\oplus\tb_p(X)\), where \(\tb_p(X) = \Z_{p_1(X)}\oplus\cdots\oplus\Z_{p_N(X)}\) is the torsion part. Then, we can obtain from \cite{Hossjer:2023ajl}, upon a slight modification to account for the multi-component case and to scale away numerical coefficients, that
\begin{equation}\label{eq:bulkPIwithtorsion}
	\parti_\BF[X] = \abs{\det\bbK}^{\h_{p}(X)}\; \abs{\hom(\tb_p(X),\Z^\kappa\big/\im\bbK)}\; \tb_\t{RS}[X]^{(-1)^{p-1}}.
\end{equation}
Note that Poincaré duality and the universal coefficient theorem give \(\tb_p(X)\cong\tb_{d-p-1}(X)\), making this expression invariant upon exchanging \(A\) and \(B\). At this stage, the inclusion of torsion seems like a mathematical curiosity. Indeed, we will not discuss further manifolds with torsion in this paper. However, this result may be of relevance for a future utilization of \textquote{surgery,}\footnote{In this method one needs to compute the partition function of BF theory on a branched cyclic cover of \(X\), ramified over the entangling surface. Although the topology of the replica manifold is completely encoded in the topology of the entangling surface and the original manifold, the replica manifold can have non-trivial torsion, even if the original manifold does not \cite{Sumners:1974homology}.} see the discussion, \Cref{sec:disc}.

\subsection{On a manifold with boundary}\label{ssec:boundaries}

Consider now the case where \(X\) has a boundary, $\pd X\neq \varnothing$. As explained in \cite{Fliss:2023dze}, the action \Cref{eq:BFact}, gives rise to a boundary symplectic form, which as it stands is consistent with fixing \(A\) as a boundary condition. In general, the action is supplemented by a boundary term, \(S_\pd\). This leads to a modified variational problem, specified by symplectic potential on the boundary, \(\boldsymbol{\Theta}_{\pd X}\). Explicitly, the variation of the total action
\(S_\t{full} = S_\BF + S_\pd\) reads:
\begin{align}
	\var{S_\t{full}} = \ip{\eom[A]}{\var{B}}_X + \ip{\eom[B]}{\var{A}}_X + \boldsymbol{\Theta}_{\pd X},
\end{align}
where \(\boldsymbol{\Theta}_{\pd X}\) can be put in Darboux form:
\begin{align}\label{eq:sympl-pot}
	\boldsymbol{\Theta}_{\pd X} = \int_{\pd X} \cmom\w\star_\pd\var{\cpos},
\end{align}
with \(\cpos\) and \(\cmom\) being functions of the boundary values of the fields, \(A_\pd \coloneqq i^*_\pd A\), \(B_\pd \coloneqq i^*_\pd B\). Having the symplectic potential in the form of \Cref{eq:sympl-pot} we have two options to proceed, in order to have a well-defined variational problem: fix \(\cmom\demeqq 0\) or \(\var\cpos\demeqq 0\). Without loss of generality, we will proceed with the Dirichlet approach, \(\cmom\demeqq 0\). The other boundary condition can be attained by a boundary symplectomorphism and can be easily shown to give equivalent results. Note that the symplectic potential is, as it stands, degenerate due to boundary gauge transformations. We will take care of that below when we discuss the boundary path integral measure.

The object one would like to study is the partition function of the combined system:
\begin{equation}\label{eq:Z-full-def}
	\parti_{\BF+\edge}[X,\pd X] \coloneqq \int_{\cC_\cmom} \frac{\DD{A}\DD{B}}{\vol(\cG)}\ \ex{-S_\t{full}^\t{E}[A,B]},
\end{equation}
where \(\cC_\cmom \coloneqq {\set{A\in \Omega^{p}(X),\ B\in \Omega^{d-p-1}(X)\suchthat \cmom = 0}}\) and \(S_\t{full}^\t{E}[A,B]\) is the Euclideanized partition function.\footnote{Note that the bulk piece is not affected by the Wick rotation, since it is a one-derivative action. However, the boundary terms can change in the usual way.} Note also that here the gauge group $\cG$ consists of gauge transformations that vanish on the boundary. The next step is to decompose the fields \(A\) and \(B\) as follows:
\begin{align}
	A & = \widetilde{A}_\pd + \hat{A}, \\
	B & = \widetilde{B}_\pd + \hat{B},
\end{align}
where \(\hat{A}\) and \(\hat{B}\) are off-shell gauge fields with Dirichlet boundary conditions, \(i^*_\pd\, \hat{A}=0=i^*_\pd\, \hat{B}\). The boundary conditions are absorbed completely by \(\widetilde{A}_\pd\) and \(\widetilde{B}_\pd\). The latter are, in the spirit of \cite{Cattaneo:2015vsa}, potentially discontinuous extensions of the boundary fields, \(A_\pd\) and \(B_\pd\) and are chosen to be on-shell, i.e. flat. This has the effect of disentangling the bulk, from the boundary contribution:
\begin{equation}
	S_\t{full}[A,B] = S_\BF\qty[\hat{A},\hat{B}] + S_\pd\qty[A_\pd,B_\pd].
\end{equation}
Similarly, the measure has a piece coming from the hatted fields and giving only bulk contributions, and one coming from the tilded fields, giving all the edge contributions.

The hatted fields are very easy to deal with, they only contribute in the bulk, and they give the full bulk contribution. The computation is completely analogous to that of \Cref{ssec:closed}. The only difference is that we make use of the Hodge decomposition theorem for Dirichlet forms on a manifold with boundary \cite{cappellCohomologyHarmonicForms2005}:
\begin{equation}
	\Omega_\t{D}^\blob(X) = \H^\blob(X,\pd X)\oplus \dd{\Omega_\t{D}^{\blob-1}(X)}\oplus\qty(\cdd{\Omega_\t{N}^{\blob+1}}\cap \Omega_\t{D}^\blob(X)),
\end{equation}
where
\begin{align}
	\Omega_\t{D}^\blob(X) & \coloneqq \set{\omega\in \Omega^\blob(X) \suchthat i_\pd^* \omega= 0}      \\
	\Omega_\t{N}^\blob(X) & \coloneqq \set{\omega\in \Omega^\blob(X) \suchthat i_\pd^* \star\omega= 0}
\end{align}
denote the spaces of Dirichlet and Neumann forms respectively. Altogether we have that
\begin{align}\label{eq:Z-bulk}
	\parti_\t{bulk}[X] = \int \frac{\DD{\hat{A}}\DD{\hat{B}}}{\vol(\cG)}\ \ex{\ii S_\BF\qty[\hat{A},\hat{B}]} = \abs{\det\bbK}^{\h_{p}\qty(X,\pd X)}\tb_\t{RS}\qty[X,\pd X]^{(-1)^{p-1}},
\end{align}
where \(\h_{p}\qty(X,\pd X)\) is the relative version of \(\h_{p}(X)\), appearing in \Cref{eq:ZBF}, i.e.:
\begin{align}\label{eq:hp-rel}
	\h_{p}(X,\pd X) & \coloneqq \Half(-1)^p\sum_{k=0}^p(-1)^k\dim \H^k(X,\pd X) + \Half(-1)^{d-p-1}\sum_{k=0}^{d-p-1}(-1)^k \dim \H^k(X,\pd X) \nn
	                & \overset{\upchi}{=} (-1)^p\sum_{k=0}^p(-1)^k\dim \H^k(X,\pd X).
\end{align}

The contribution of the tilded fields is a little more subtle. We will calculate their contribution, with a specific choice of \(S_\pd\), in detail in \Cref{sec:replica}, but let us already give a rough sketch of the computation for a general boundary action. First of all, it is clear from the above discussion that they only contribute on the boundary. Secondly, the flatness constraint becomes a Bianchi identity for the boundary gauge fields. In particular, $A_\pd$ and $B_\pd$ are curvatures of $(p-1)-$ and $(d-p-1)$-form gauge fields, \(a\) and \(b\), respectively. All in all, the contribution of $\widetilde{A}_\pd$ and $\widetilde{B}_\pd$, including the boundary condition $\cmom=0$, in the measure is
\begin{equation}\label{eq:bdy-measure}
	\sum_{\t{instantons}} \int \frac{\DD{a}\DD{b}}{\vol(\cG^\pd)}\ \delta[\cmom],
\end{equation}
where the sum over instantons is a sum over all topologically non-trivial configurations, i.e. over cohomology \(p\)- and $(d-p-1)$-classes, respectively, of $\pd X$, and $\vol(\cG^\pd)$ indicates the volume of the boundary gauge transformations. Since \(a\) and \(b\) appear only through their curvatures, there is a redundancy upon shifting them by flat gauge fields. Relatedly, dividing by this volume takes care of the redundancy in the symplectic potential, $\boldsymbol{\Theta}_{\pd X}$.

Putting everything together, we find that the partition function, \Cref{eq:Z-full-def}, takes the form
\begin{align}\label{eq:Z-bulk+edge}
	\parti_{\BF+\edge}[X, \pd X] =\parti_\t{bulk}[X] \parti_{\edge}[\pd X],
\end{align}
with $\parti_\t{bulk}[X]$ as in \Cref{eq:Z-bulk} and $\parti_{\edge}[\pd X]$ is the edge-mode partition function, computed with the measure \Cref{eq:bdy-measure}.

\section{Entanglement from the replica path integral}\label{sec:replica}

In this section, we will calculate the entanglement entropy using path integral techniques via the replica trick. We begin with a brief primer on replica path integrals and how they are calculated. The basic ingredient is the path integral representation of a state. We begin with a Cauchy slice, $\Sigma$, and a Hilbert space of states associated to it, $\cH_\Sigma$. We can imagine generating a state in $\cH_\Sigma$ by finding a \(d\)-dimensional manifold, $X_-$, whose boundary is $\Sigma$. According to the standard rules of topological field theory, the path integral with specified boundary conditions produces the wavefunction of a state in $\cH_\Sigma$. That is,
\begin{equation}
	\Psi_{X_-}[\varphi]\coloneqq \int_{\cC[X_-;\varphi]} \DD\Phi\ \ex{\ii S[\Phi]},
\end{equation}
is the wavefunction associated with the state
\begin{align}
	\ket{\Psi_{X_-}} = \int_{\cC[\Sigma]}\DD{\varphi} \Psi_{X_-}[\varphi]\ket{\varphi}\in\cH_\Sigma,
\end{align}
in the Hilbert space \(\cH_\Sigma\) assigned to \(\Sigma\). In the above, we have schematically indicated all fields by $\Phi$ and \(\cC[X_-;\varphi]\) is an appropriate functional space (including quotienting out gauge redundancies) over \(X_-\), with boundary conditions \(i_{\pd X_-}^*\Phi=\varphi\). \(\cC[\Sigma]\) is, similarly, an appropriate functional space over \(\Sigma\). The dual Hilbert space, \(\chk{\cH_\Sigma}\) is canonically isomorphic with the Hilbert space associated to the orientation reversal, \(\overline{\Sigma}\), of \(\Sigma \). Hence, the norm of the state \(\ket{\Psi_{X_-}}\), is given by conjugating $\Psi_{X_-}$ and integrating over the boundary conditions. At the level of the path integral, this has the action of gluing $X_-$ onto $X_+$, its time-reversal about $\Sigma$,
\begin{equation}
	\norm{\Psi_{X_-}}^2=\int_{\cC[X]}\DD\Phi\ \ex{\ii S[\Phi]} = \parti[X], \label{eq:norm-Psi}
\end{equation}
yielding the partition function on $X\coloneqq X_-\cup_{\Sigma}X_+$. Similarly, the pure density matrix \linebreak \(\rho \coloneqq \ketbra{\Psi_{X_-}}\in\cH_\Sigma\otimes\chk{\cH_\Sigma}\) has the path integral expression
\begin{align}
	\rho[\varphi_-,\varphi_+] & \coloneqq \mel{\varphi_-}{\rho}{\varphi_+} = \Psi_{X_-}[\varphi_-] \Psi_{X_-}^*[\varphi_+] = \nn
	                          & = \int_{\cC[X_{\Sigma\t{-cut}};\varphi_-,\varphi_+]} \DD{\Phi}\ \ex{\ii S[\Phi]},
\end{align}
where \(X_{\Sigma\t{-cut}} \coloneqq X_-\sqcup X_+ \cong X \setminus \qty(\Sigma \times [0,1])\), which has as its boundary two copies of $\Sigma$ with opposite orientation, \(\Sigma_\pm\), on which we impose boundary conditions \(\varphi_\pm\), respectively. This density matrix could possibly be unnormalized.

We now choose a subregion $\aent\subset\Sigma$ and imagine demarcating
\begin{equation*}
	\varphi_\pm=\Theta_\aent\,\varphi_\pm+\Theta_{\coaent}\,\varphi_\pm \eqqcolon \varphi_{\aent,\pm}+\varphi_{\coaent,\pm},
\end{equation*}
where $\Theta_\aent$ is the characteristic function of $\aent$ (taking values 1 inside $\aent$ and 0 everywhere else) and $\Theta_{\coaent}=1-\Theta_\aent$ is the characteristic function for the complement region. The reduced density matrix, $\rho_\aent$, is described, at least at a formal level, by identifying $\varphi_{\coaent,+}=\varphi_{\coaent,-}$ and integrating over their values:
\begin{align}
	\rho_\aent[\varphi_{\aent,-},\varphi_{\aent,+}] & = \int_{\cC[\coaent]}\DD\varphi_{\coaent}\rho[\varphi_{\aent,-}+\varphi_{\coaent};\overline{\varphi}_{\aent,+}+\varphi_{\coaent}]=\int_{\cC[X_{\aent\t{-cut}};\varphi_{\aent,-},\varphi_{\aent,+}]}\DD\Phi\ \ex{\ii S[\phi]},
\end{align}
resulting in a path integral on $X_{\aent\t{-cut}}\coloneqq X_-\cup_{\coaent}X_+$, possessing a cut along $\aent$, and boundary conditions $\varphi_\pm$, imposed on either side of the cut. This density matrix is unnormalized: its trace is simply $\norm{\Psi_{X_-}}^2=\parti[X]$, which arises from identifying $\varphi_{\aent,+}$ with $\varphi_{\aent,-}$ and then integrating, effectively \textquote{gluing} the cut closed.

We now aim to calculate the von Neumann entropy of $\rho_\aent$. We will make use of the replica trick.  First we compute the \(n\)-th R\'enyi entropy as
\begin{equation}\label{eq:Renyidef}
	\cS_n\coloneqq\frac{1}{1-n}\log\frac{\Tr\left(\rho_\aent^n\right)}{\left(\Tr\rho_\aent\right)^n}~,
\end{equation}
where the denominator arises to normalize $\rho_\aent$, if it was not already normalized.  The von Neumann entropy is then given as the limit:
\begin{equation}
	\ee=\lim_{n\rightarrow 1}\cS_n~.
\end{equation}
The denominator of $\Cref{eq:Renyidef}$ is simply given by $\parti[X]^n$, so we now make sense of the numerator.  It arises from copying $X_{R\t{-cut}}$ $n$ times; the trace identifies the boundary conditions, $\varphi^{(i)}_{\aent,+}=\varphi^{(i+1)}_{\aent,-}$ (where \(i\) indexes the replicas mod \(n\)) and their subsequent integration glues the replicated manifold into an \(n\)-fold branched cover over $\pa\aent$, which we denote $X_{n}$. This gives a path integral expression of the entanglement entropy as
\begin{equation}\label{eq:EEasRepPIs}
	\ee=\lim_{n\rightarrow 1}\frac{1}{1-n}\log\frac{\cZ[X_{n}]}{\cZ[X]^n} = -\eval{\pdv{n}(\log\frac{\parti[X_n]}{\parti[X]^n})}_{n=1}.
\end{equation}
We are now tasked with evaluating the path integrals in question taking care with the codimension-2 surface fixed by the $\mathbb Z_n$ replica symmetry --- the entangling surface. We will address this by regulating the entangling surface and looking at the edge theory that arises there. To be precise, we will literally excise a tubular neighborhood around the entangling surface: in terms of the state on $\Sigma$, this has the interpretation of putting a small buffer region between $\aent$ and $\coaent$.  This will serve as a UV regulator.  Specifically, we replace the bulk replica manifold $X_{n}$ with $\rep{n}\coloneqq X_{n}\setminus \left(\disc^2_{n\varepsilon}\times\pa\aent\right)$ and keep in mind the limit as $\varepsilon\rightarrow0$. The interpretation of $\varepsilon$ as a regulator on the unreplicated state mandates that the circumference of the disk scales as $n\varepsilon$. We will perform a similar excision of the original manifold: $X\leadsto\rep{1}\coloneqq X\setminus\left(\disc^2_{\varepsilon}\times\pa\aent\right)$. This yields
\begin{equation}\label{eq:SEE_as_limit_of_ratio}
	\ee=\lim_{\varepsilon\rightarrow0}\lim_{n\rightarrow1}\frac{1}{1-n}\log\frac{\parti_{\BF+\edge}\qty[\rep{n}]}{\parti_{\BF+\edge}\qty[\rep{1}]^n}~.
\end{equation}
Importantly, these are path integrals on manifolds with boundary, $\brep{n}=\S^1_{n \varepsilon}\times\pd\aent$. As we have discussed in \Cref{sec:BF-PI}, we need to supplement these path integrals with boundary conditions and boundary terms on $\rep{n}$ enforcing those boundary conditions through the variational principle. As we have shown in \Cref{ssec:boundaries}, the BF path integral on this regulated, replica geometry naturally splits into a product of \textquote{bulk} and \textquote{edge} terms
\begin{equation}
    \parti_{\BF+\edge}\qty[\rep{n},\brep{n}]=\parti_{\text{bulk}}\qty[\rep{n}]\;\parti_\edge\qty[\brep{n}].
\end{equation}
We now insert the relevant $\parti_{\BF+\edge}\qty[\rep{n},\brep{n}]$ into the ratio \Cref{eq:SEE_as_limit_of_ratio}. In this paper, we will focus on the contribution of the edge-modes to the entanglement entropy. It is precisely these terms that we expect to contribute an area law. In the discussion, \Cref{sec:disc}, we will comment on possible bulk contributions. Since the above appears as a product, we can neatly isolate the contribution of the edge-modes as:
\begin{align}\label{eq:EE-replica}
	\ee = \lim_{\varepsilon\to 0}\lim_{n\to 1} \frac{1}{1-n}\log\frac{\parti_\edge\qty[\brep{n}]}{\parti_\edge\qty[\brep{1}]^{n}}.
\end{align}

\subsection{The edge-mode theory}\label{ssec:edge-theory}

We now consider the boundary action. The bare BF action, \Cref{eq:BFact}, is consistent with either fixing \(A\) or annihilating \(B\) as a boundary condition. However, the subsequent integration over the remaining fluctuating boundary degrees of freedom is now unbounded: we must supplement this theory with a boundary Hamiltonian.\footnote{The need to supplement an edge Hamiltonian as a regulator is a theme that will be echoed in \Cref{sect:edgemodes}.} We will focus on quadratic boundary actions. In general dimensions these quadratic actions will be dimensionful and general lessons of effective field theory guide us to write the most relevant one.  Without loss of generality we can assume that this comes from the quadratic action for the \(p\)-forms and write the following boundary action:
\begin{equation}\label{eq:bdy-action}
	S_\pd[A_\pd,B_\pd] = \frac{\bbK}{4\pi}\mu^\Delta \int_{\brep{n}} A_{\pd} \w\star A_\pd = \frac{\mu^\Delta}{4\pi}\ip{A_\pd}{\bbK\, A_\pd},
\end{equation}
where \(\mu\) is an arbitrary energy scale and $\Delta = d-1-2p$, as required by dimensional analysis. With this action, the boundary symplectic form (taking also into account the boundary terms from the variation of the bulk action) becomes:
\begin{equation}
	\boldsymbol{\Theta}_{\brep{n}} = \int_{\brep{n}} \qty(B_\pd + \mu^\Delta (-1)^{(d-p-1)(p+1)} \star A_\pd)\w \var{A_\pd}.
\end{equation}
As such, the variational principle, with the above boundary action, \Cref{eq:bdy-action}, is compatible with demanding
\begin{equation}\label{eq:chiral-bc}
	B_\pd + \mu^\Delta (-1)^{(d-p-1)(p+1)} \star A_\pd \demeqq 0
\end{equation}
as boundary condition. We will call \Cref{eq:chiral-bc} a \textquote{generalized chiral boundary condition.}

This boundary action, \Cref{eq:bdy-action}, and the corresponding boundary condition, \Cref{eq:chiral-bc}, have been known for a long time to appear on the boundary of BF theories,  \cite{Maldacena:2001ss, Kravec:2013pua}. Ignoring global issues, the flatness condition that the bulk path integral imposes on \(A_\pd\) makes this is a simple theory of a free \((p-1)\)-form gauge field, also known as a singleton mode in the string theory literature. Naturally (Hodge-dually), it is also a theory of a free \((d-p-2)\)-form. However, on a non-trivial topology, the boundary condition \Cref{eq:chiral-bc} imposes a constraint on the fluxes. On a manifold of the form \(\S^1_{\varepsilon}\times \Sigma\) --- as is our case here --- both fields have fluxes obeying Dirac quantization around purely spatial cycles, but fractional charges, when wrapping the thermal circle. This is entirely reminiscent of the story of the two-dimensional chiral boson on a torus, which can be seen to arise at the edge of a Chern--Simons theory, which is \emph{not} modular invariant: upon demanding periodicity around the spatial circle, one cannot retain periodicity around the thermal circle. It also reflects the story of self-dual gauge fields in higher dimensions \cite{freedUncertaintyFluxes2007,freedHeisenbergGroupsNoncommutative2007}.\footnote{Of course our approach is much less rigorous than the original story. A rigorous analysis would require a revisit from the lens of differential cohomology. We take this as an opportunity to stress its importance for a solid understanding of gauge theories.}  In our case, the theory becomes a \textquote{chiral half} of a \((p-1)\)-form and a \((d-p-2)\)-form Maxwell theory, which we call, relatively unimaginatively, \emph{chiral mixed Maxwell theory}. The role of chiral fields as edge-modes of BF theories was recently reemphasized in \cite{Arvanitakis:2022bnr,Fuentealba:2023usq,Evnin:2023cdf}. Note, however, that in most instances where chiral gauge fields make an appearance (see e.g. \cite{Pasti:1995ii,freedHeisenbergGroupsNoncommutative2007,freedUncertaintyFluxes2007,henningsonQuantumHilbertSpace2002,Pasti:1996vs,Perry:1996mk,Pasti:2012wv,Szabo:2012hc,Witten:1996hc,Witten:1999vg,Belov:2006jd,Andriolo:2021gen,Avetisyan:2021heg,Avetisyan:2022zza} for an incomplete list) the story is focussed on \(k\)-form fields in \(2k+2\) dimensions, where one can construct a genuinely chiral combination of the gauge fields. This is, to our knowledge, the first time a chiral pure gauge theory is treated quantum mechanically in generic dimensions.\footnote{And as a bonus, arbitrary topology.}\textsuperscript{,}\footnote{At the classical level this theory is considered in \cite{Evnin:2023cdf}, where our generalized chiral boundary condition goes under the moniker twisted self-duality. We thank the authors of \cite{Evnin:2023cdf} for discussions regarding this point.}

Let us be more concrete. First, we Hodge-decompose \(A_\pd\) as \(A_\t{h}+\dd{a}\), where \(A_\t{h}\) is a harmonic \(p\)-form and \(a\) is the (globally-well defined) \((p-1)\)-form gauge field. We do the same for \(B_\pd\). Now, if we denote by \(\dd{\tau}\) the volume element of the thermal circle, we can split the harmonic parts of the two fields as 
\begin{align}
    A_\t{h} &= \cA + \dd{\tau} \w \widetilde{\cA} \\
    B_\t{h} &= \cB + \dd{\tau} \w \widetilde{\cB},
\end{align}
where \(\cA\in \harm^p(\pd\aent)\), \(\widetilde{\cA}\in \harm^{p-1}(\pd\aent)\), \(\cB\in\harm^{d-p-1}(\pd\aent)\), and \(\widetilde{\cB}\in\harm^{d-p-2}(\pd\aent)\). Then the boundary condition \Cref{eq:chiral-bc} relates \(\widetilde{\cA}\) to \(\cB\) (and similarly \(\widetilde{\cB}\) to \(\cA\)) as follows:
\begin{equation}
    \cB = \mu^\Delta (-1)^{(d-p)p}\star_{\pd\aent} \widetilde{\cA}.
\end{equation}
And the boundary action becomes
\begin{align}\label{eq:chiral-action}
    S_\pd\qty[\cA,\cB] = \frac{\mu^\Delta}{4\pi}\ip{\cA}{\bbK\, \cA} + \frac{\mu^{-\Delta}}{4\pi}\ip{\cB}{\bbK\, \cB} + \frac{\mu^\Delta}{4\pi}\ip{\dd{a}}{\bbK\, \dd{a}}.
\end{align}
We take this as the definition of the action of chiral mixed Maxwell theory. 

Let us proceed with quantizing this theory, by performing the path integral on $\brep{n}$. For notational simplicity we will perform the path integral on $\brep{1}$ and we will rescale the radius in the final formulas to obtain the partition function on $\brep{n}$. First, observe that writing the action in the form \Cref{eq:chiral-action} is equivalent to performing the path integral over \(\widetilde{\cA}\) and \(\widetilde{\cB}\) over the delta function that enforces the boundary conditions. This leaves us with
\begin{equation}\label{eq:parti-edge2}
    \parti_\t{edge}\qty[\brep{1}] = \parti_\t{inst}\qty[\brep{1}]\, \parti_\t{osc}\qty[\brep{1}],
\end{equation}
where
\begin{align}
    \parti_\t{inst}\qty[\brep{1}] &\coloneqq \int \DD{\cA}\DD{\cB} \exp(-\frac{\mu^\Delta}{4\pi}\ip{\cA}{\bbK\, \cA} + \frac{\mu^{-\Delta}}{4\pi}\ip{\cB}{\bbK\, \cB}) \label{eq:parti-inst} \\
    \parti_\t{osc}\qty[\brep{1}] &\coloneqq \int \frac{\DD{a}}{\vol(\cG^\pd_{p-1})} \exp(-\frac{\mu^\Delta}{4\pi}\ip{\dd{a}}{\bbK\, \dd{a}}).
\end{align}
The instanton integral is over the space of harmonic \(p\)- and \((d-p-1)\)-forms on \(\pd\aent\) and we will evaluate it shortly. The oscillator piece is an integral over a \((p-1)\)-form gauge fields modulo their gauge transformations. This is a regular set of \((p-1)\)-form Maxwell oscillators.

We first deal with the oscillators. We can expand in modes of the Hodge Laplacian and integrate over those, separating the zero-modes. We must do so for the tower of ghosts that follow from the reducible gauge invariances of the \((p-1)\)-form gauge fields. We show in \Cref{app:parties}, that the oscillators contribute as
\begin{align}\label{eq:parti-osc}
    \parti_\t{osc}\qty[\brep{1}] = \qty(\upeta^{(p-1)}_{\pd\aent}\qty[q])^{-\kappa}\;\qty(\upeta^{(d-p-2)}_{\pd\aent}\qty[q])^{-\kappa}.
\end{align}
In the above, \(q\) denotes the nome, \(q\coloneqq \ex{-\varepsilon\mu}\), and we have defined an analogue of the Dedekind eta function,
\begin{align}\label{eq:eta-def}
    \upeta^{(k)}_{Y}[q] \coloneqq q^{-\half E_0}\qty(\prod_{\sfn\in\sN^\perp_{k}} \sum_{N_\sfn = 0}^\infty q^{N_\sfn \sqrt{\lambda_\sfn}})^{-1/2},
\end{align}
associated to the \(k\)-form Hodge Laplacian on a closed, compact manifold \(Y\). In the above, \(\sN_k^\perp\) denotes the index-set of the non-zero spectrum of the \(k\)-form Laplacian acting on coclosed forms,  \(\lambda_\sfn\) are the corresponding eigenvalues, measured in units of $\mu$, and \(E_0\) is the (potentially divergent) zero-point energy (also in units of $\mu$), \(E_0 \coloneqq \sum_{\sfn\in\sN^\perp_k} \lambda_\sfn\). Moreover, the above is, strictly speaking, defined only up to a phase. This will not concern us; the existence or choice of a complex structure of \(Y\) is insignificant for us, since for our purposes it suffices to only consider real nomes. The definition \Cref{eq:eta-def} is also equivalent to 
\begin{align}\label{eq:eta-def2}
    \upeta_{Y}^{(k)}[q] = \qty(\prod_{\sfn_k\in\sN_k^\perp} \sinh(\Half\varepsilon\mu\sqrt{\lambda_{\sfn_k}}))^{1/2} = \qty(\prod_{\sfn_k\in\sN_k^\perp} q^{-\Half\sqrt{\lambda_{\sfn_k}}} \qty(1-q^{\sqrt{\lambda_{\sfn_k}}}))^{1/2},
\end{align}
Finally it is easy to see that if we take \(Y=\S^1\), our generalized eta function reduces to the usual Dedekind eta function: \(\upeta_{\S^1}^{(0)}[q]=\upeta[q]\). Moreover, since the non-zero spectra of the Laplacians acting on transversal \((p-1)\)- and \((d-p-2)\)-forms coincide, the eta functions associated with those are equal, we write them separately, however for reasons that will become apparent shortly.

Moving on to the instantons, we can further expand \(\cA\) and \(\cB\) in terms of the topological basis of harmonic forms on \(\pd\aent\):
\begin{equation}
    \cA_\t{h} = \cA_\sfi \tau_\sfi^{(p)} \qq{and} \cB_\t{h} = \cB_\sfi \tau_\sfi^{(d-p-1)}.
\end{equation}
Enter Dirac quantization. Magnetic fluxes along \(p\)- and \((d-p-1)\)-cycles on \(\pd\aent\) are quantized as
\begin{align}
    \int_{\eta} \cA \in  2\pi\Z^\kappa, \quad \fall \eta\in\H_p(\pd\aent) \qq{and}
    \int_{\gamma} \cB \in  2\pi\Z^\kappa, \quad \fall \gamma\in\H_{d-p-1}(\pd\aent).
\end{align}
This means that \(\cA_\sfi=2\pi n_\sfi\) and \(\cB_\sfi = 2\pi m_\sfi\), where \(n_\sfi,m_\sfi\in\Z^\kappa\). As such, the action \Cref{eq:chiral-action} becomes:
\begin{align}
    S_\pd\qty[\cA,\cB] &= \pi \varepsilon \mu\qty(\vec{n}\cdot\qty(\bbK\otimes\widetilde{\bbG}_p^{\pd\aent})\cdot\vec{n} + \vec{m}\cdot\qty(\bbK\otimes\widetilde{\bbG}_{d-p-1}^{\pd\aent})\cdot\vec{m}).
\end{align}
In the above, we have combined \(n_\sfi\) and \(m_\sfi\) into vectors \(\vec{n}\in\Z^{\kappa\,\b_p(\pd\aent)}\) and \(\vec{m}\in\Z^{\kappa\,\b_{d-p-1}(\pd\aent)}\). Moreover, \(\widetilde{\bbG}_k^{\pd\aent}\) is the dimensionless version of \(\bbG_k^{\pd\aent}\), the Gram matrix of harmonic \(k\)-forms on \(\pd\aent\). This comes from the fact that \(\bbG_k\) has dimension \(\qty[\bbG_k] = \mu^{1+(2k-1-d)}\), so \(\widetilde{G}_p \coloneqq \mu^{\Delta-1}\) and \(\widetilde{\bbG}_{d-p-1} \coloneqq \mu^{1+\Delta}\bbG_{d-p-1}\) are dimensionless. Therefore, the instanton contribution reads
\begin{equation}
    \parti_\t{inst}\qty[\brep{1}] = \sum_{\substack{\vec{n}\in\Z^{\kappa\b_p(\pd\aent)}\\ \vec{m}\in\Z^{\kappa\b_{d-p-1}(\pd\aent)}}} \hspace{-1em}\exp[-\pi \varepsilon \mu\qty(\vec{n}\cdot\qty(\bbK\otimes\widetilde{\bbG}_p^{\pd\aent})\cdot\vec{n} + \vec{m}\cdot\qty(\bbK\otimes\widetilde{\bbG}_{d-p-1}^{\pd\aent})\cdot\vec{m})]. \label{eq:parti-inst-nm} 
\end{equation}

To arrive at the final expression for the edge-mode partition function, let us also introduce a Siegel-type Theta function, denoted by \(\Theta[q;\bbO_N]\):
\begin{align}\label{eq:Theta}
    \Theta[q;\bbO_N] \coloneqq \sum_{\vec{r}\in \Z^N} q^{\pi \vec{r}\cdot\bbO_N\cdot\vec{r}}.
\end{align}
If \(\bbO_N\) is a real, \(N\times N\) matrix, when \(q\in\R_{\geq 0}\), it truly becomes a Siegel Theta function and inherits all the modular properties of those. Expressing the instanton contribution, \Cref{eq:parti-inst}, in terms of the Theta functions, \Cref{eq:Theta}, and combining with the oscillator contribution, \Cref{eq:parti-osc}, gives the following simple form for the edge-mode partition function:
\begin{empheq}[box=\obox]{equation}\label{eq:parti-edge-1}
    \parti_\t{edge}\qty[\brep{1}] = \frac{\Theta\qty[q;\bbK\otimes\widetilde{\bbG}_p]}{\qty(\upeta^{(p-1)}_{\pd\aent}[q])^\kappa}\ \frac{\Theta\qty[q;\bbK\otimes\widetilde{\bbG}_{d-p-1}]}{\qty(\upeta^{(d-p-2)}_{\pd\aent}[q])^\kappa}.
\end{empheq}

We briefly re-emphasize the chiral nature of this result and comment on its importance. To get a feel for it, note first that if \(\pd\aent=\S^1\), with \(p=1\), this becomes exactly the partition function of two (multi-component) chiral bosons. This resonates entirely with the usual story. In \(d=3\) BF theory can be written as a sum of two Chern--Simons theories, each of which supports a chiral scalar on its boundary. The partition function of this chiral boson is {\it not} modular invariant\footnote{This fact, $\parti_\text{chiral boson}(\beta)\neq\parti_\text{chiral boson}(1/\beta)$, is actually crucial for extracting the correct entanglement entropy of bulk Chern-Simons theory \cite{Wen:2016snr}!}. In \(d=5\) dimensions, a smoking gun signal that the edge theory is chiral is the fact that it is not \(\sfS\)-duality invariant, as can be seen by comparing to the regular Maxwell partition function (cf. \Cref{app:parties}). In \cite{Kravec:2013pua}, \(\sfS\)-duality of the edge-mode theory was crucial in proving an obstruction to symmetry-preserving regulators of edge-states of BF theory.\footnote{This was, in turn interpreted as a gauge theory version of the Nielsen--Ninomiya theorem \cite{Nielsen:1980rz,Nielsen:1981xu}.} Our analysis provides evidence that this may not be the case; i.e. the fact that the UV lattice models regularizing Maxwell theory break \(\sfS\)-duality, while correct, does not imply that these models are symmetry breaking from the point of view of edge states of BF theory.  Moving away from \(d=5\), for generic dimensions, one can take a half of our partition function as the definition of an Abelian chiral gauge theory, but note that only this combination is well-defined, at least from a path integral perspective. A systematic understanding of this chiral theory deserves further research.

In the next section, we will move on to extract the entanglement entropy. To do that, we will need to compare the edge-mode partition function on \(\brep{n}\) to that on \(\brep{1}\). The \(\brep{n}\) partition function can be easily obtained by rescaling the radius of the \(\S^1\) to \(n\varepsilon\). This results in
\begin{equation}\label{eq:parti-edge-n}
    \parti_\t{edge}\qty[\brep{n}] = \frac{\Theta\qty[q^n;\bbK\otimes\widetilde{\bbG}_p]}{\qty(\upeta^{(p-1)}_{\pd\aent}\qty[q^n])^\kappa}\ \frac{\Theta\qty[q^n;\bbK\otimes\widetilde{\bbG}_{d-p-1}]}{\qty(\upeta^{(d-p-2)}_{\pd\aent}\qty[q^n])^\kappa}.
\end{equation} 

\subsection{Extracting the entanglement entropy}\label{sec:edgethyentanglementcalc}

We can now simply insert \Cref{eq:parti-edge-n,eq:parti-edge-1} into \Cref{eq:EE-replica} to extract the entanglement entropy. We have two kinds of terms to deal with; the Theta functions and the eta functions. Let us deal with them separately.

We treat first the instantons, i.e. the Theta functions, for it will be easier to extract their high-temperature behavior. Fortunately, Siegel Theta functions are very well behaved and obey a sort of modularity equation, which can be easily proved using the Poisson summation formula. Explicitly, they obey the following transformation equation \cite{TataLecturesTheta2007}:
\begin{align}
	\Theta[\ex{-\varepsilon\mu};\bbO_N] = \qty(\varepsilon\mu)^{-\frac{N}{2}} \abs{\det\bbO_N}^{-\frac{1}{2}} \Theta\qty[\ex{-\frac{1}{\varepsilon\mu}};\bbO_N^{-1}].
\end{align}
The high-temperature behavior is then straightforward to obtain:
\begin{align}
	\Theta[\ex{-\varepsilon\mu};\bbO_N] \overset{\varepsilon\to 0}{\sim} \qty(\varepsilon\mu)^{-\frac{N}{2}} \abs{\det\bbO_N}^{-\frac{1}{2}}.
\end{align}
With that, the total contribution of the Theta functions to the entanglement entropy is
\begin{align}\label{eq:ee-thetas}
	\eval{\ee}_{\Theta} & = \lim_{\varepsilon\to 0}\lim_{n\to 1}\frac{1}{1-n}\log\frac{\Theta\qty[q^n;\bbK\otimes\widetilde{\bbG}^{\pd\aent}_p]\ \Theta\qty[q^n;\bbK\otimes\widetilde{\bbG}^{\pd\aent}_{d-p-1}]}{\qty(\Theta\qty[q;\bbK\otimes\widetilde{\bbG}^{\pd\aent}_p]\ \Theta\qty[q;\bbK\otimes\widetilde{\bbG}^{\pd\aent}_{d-p-1}])^n} = \nonumber \\[0.4em]
	                    & = \frac{1}{2}\qty(\b_p(\pd\aent)+\b_{d-p-1}(\pd\aent))\qty\big[\kappa\qty(1-\log(\varepsilon \mu))-\log\abs{\det\bbK}].
\end{align}
As we shall see shortly, this term will be partially cancelled by the contribution of the eta functions, ultimately resulting in the topological piece of the entanglement entropy.

Let us get to it then. To evaluate the contribution of the eta functions, we will use the following trick \cite{debruyne2020saddle,Fliss:2021ekk}.  In words, we will Taylor expand $\log\qty(\upeta_{\pa\aent})^{-1}$ and on each term perform an inverse Mellin tranform\footnote{Note that the piece containing the zero-point-energy, $E_0$, in \Cref{eq:eta-def} or \Cref{eq:eta-def2} obviously does not contribute to the entropy so we will focus on the rest.}.  Written as a Mellin integral, the sum over the eigenvalues, $\set{\lambda_\sfn}$ is easily performed to yield the spectral zeta function,
\begin{align}\label{eq:speczeta}
	\sz{\pd\aent}{k}(s)\coloneqq\sum_{\sfn\in\sN^\perp_{k}}\lambda_\sfn^{-s}.
\end{align}
In equations:
\begin{align}\label{eq:desc-manipulations}
	- \sum_{\sfn\in\sN_k^\perp} \log(1-\ex{-\varepsilon\mu\sqrt{\lambda_\sfn}})^{-1} & \overset{\t{Taylor}}{=\joinrel=\joinrel=}  \sum_{\sfn\in\sN_k^\perp}\sum_{m=1}^\infty\frac{1}{m} \ex{-\varepsilon\mu\;  m \sqrt{\lambda_\sfn}}\nonumber                                                         \\
	                                                                           & \overset{\inv{\t{Mellin}}}{=\joinrel=\joinrel=} \sum_{\sfn\in\sN_k^\perp}\sum_{m=1}^\infty\frac{1}{m}\int_{c+\ii\R}\frac{\dd{u}}{2\pi\ii} \Gamma(u) \qty(\varepsilon\mu\; m \sqrt{\lambda_\sfn})^{-u} \nonumber \\
	                                                                           & \overset{\t{resum}}{=\joinrel=\joinrel=} \int_{c+\ii\R}\frac{\dd{u}}{2\pi\ii}\; \upzeta(u+1)\; \sz{\pd\aent}{p-1}\qty(\frac{u}{2})\; \Gamma(u)\; \qty(\varepsilon\mu)^{-u}  ,
\end{align}
where $\upzeta(u)$ is the ordinary Riemann zeta function.  In the above integral \(c>0\) is a positive number lying to the right of the rightmost pole of the integrand. To extract the leading terms as $\varepsilon\rightarrow0$, we then imagine pushing the integration contour to \(-1<c<0\), so that the integrand vanishes as \(\varepsilon\to 0\). The price we pay is that, as we sweep to the left, we pick up the residues from all of the poles of the integrand.
\begin{align}\label{eq:limlogasres}
	\lim_{\varepsilon\rightarrow0}\log\qty(\upeta^{(k)}_{\pa\aent}(q))^{-1}=\Half\sum_{u_*\geq0}\underset{u=u_*}{\Res} \upzeta(u+1)\; \sz{\pd\aent}{k}\qty(\frac{u}{2})\; \Gamma(u)\; \qty(\varepsilon\mu)^{-u}.
\end{align}

In short, evaluating the descendant contribution to $\ee$ reduces to analyzing the pole structure of the above expression.  In \Cref{app:zeta} we carefully analyze the poles of $\sz{\pd\aent}{k}$ which all lie at $u>0$. Inside \Cref{eq:limlogasres} there is a double pole at \(u=0\) from \(\Gamma(u)\upzeta(u+1)\), with residue
\begin{align}
	\Res_0^{(k)} & = \Half{\sz{\pd\aent}{k}}'(0) - \sz{\pd\aent}{k}(0)\log\qty(\varepsilon\mu) \nn
	                & =\qty(-\cI_{\frac{d-2}{2}}^{(k)} \delta_{d,\t{even}} + \b_k(\pd\aent))\log\qty(\varepsilon\mu).
\end{align}
In the above, \(\cI_{\frac{d-2}{2}}^{(k)}\) is the \(\qty(\frac{d-2}{2})\)-th heat kernel coefficient for the spectral zeta function arising from the value of \(\sz{\pd\aent}{k}(0)\) and $\b_k(\pd\aent)$, the \(k\)-th Betti number enumerates the number of zero-modes of the $k$-form Laplacian, again reviewed in \Cref{app:zeta}. We see then that this term, evaluated for the eta functions associated to the $(p-1)$- and the $(d-p-2)$-form Laplacians, contributes to the entropy as:
\begin{align}\label{eq:ee-Res0}
	\eval{\ee}_{\Res_0} = \frac{\kappa}{2} \qty[\qty(\cI_{\frac{d-2}{2}}^{(p-1)}+\cI_{\frac{d-2}{2}}^{(d-p-2)}) \delta_{d,\t{even}} - \qty(\b_{p-1}(\pd\aent)+\b_{d-p-2}(\pd\aent))]\qty(1-\log\qty(\varepsilon\mu)).
\end{align}
We see a happy cancellation between (part of) this term and (part of) the contribution of the Thetas, \Cref{eq:ee-thetas}, owing to Poincaré duality \(\b_{p-1}(\pd\aent)=\b_{d-p-1}(\pd\aent)\) and \(\b_{d-p-2}(\pd\aent)=\b_p(\pd\aent)\).

The residues away from zero come from the poles of \(\sz{\pd\aent}{p-1}(u/2)\).\footnote{Recall that apart from the zero-modes, the eta functions for $(p-1)$- and for $(d-p-2)$-forms are identical.} These lie at \(d-2k\), for integer \(k\geq 1\)\footnote{Note that there is a shift \(k\to k-1\) compared with \Cref{app:zeta}.} (with \(k\leq d/2-1\) when \(d\) is even) and give in total
\begin{align}
	\Res_{2d-k} = (4\pi)^{1-\frac{d}{2}} \upzeta(d-2k+1) \frac{\Gamma(d-2k)}{\Gamma\qty(\frac{d}{2}-k)} \cI_{k-1}^{(p-1)} \qty(\varepsilon\mu)^{2k-d},
\end{align}
where, again, \(\cI_{k-1}^{(p-1)}\) are heat kernel coefficients for the spectral zeta function. They can be written as integrals of local geometrical (but not topological) data of the entangling surface. These residues contribute area law (and subleading) terms to the entropy:
\begin{align}
	\eval{\ee}_{\Res_{d-2k}} & = \mathrm{I}^{(p-1)}_k \qty(\varepsilon\mu)^{2k-d},
\end{align}
where
\begin{align}\label{eq:A-coeffs}
	\mathrm{I}^{(p-1)}_k & \coloneqq \kappa (4\pi)^{1-\frac{d}{2}} \upzeta(d-2k+1) \frac{\Gamma(d-2k)}{\Gamma\qty(\frac{d}{2}-k)} (d-2k+1)\ \cI_{k-1}^{(p-1)}.
\end{align}
As expected, the area-law terms are non-universal, depend on the geometry of the entangling surface (through the heat-kernel coefficients, $\cI_{k-1}^{(p-1)}$) and the regulator, $\varepsilon$.

Assembling the oscillator contributions and adding the contribution of the instantons, we arrive at the main result for the entanglement entropy:
\begin{equation}\label{eq:EE-replica-full}
	\begin{aligned}
		\ee & = \sum_{k=1}^{\left\lfloor\frac{d-1}{2}\right\rfloor} \mathrm{C}^{(p-1)}_k \qty(\frac{\ell}{\varepsilon})^{d-2k} + \frac{\kappa}{2} \qty(\cI_{\frac{d-2}{2}}^{(p-1)}+\cI_{\frac{d-2}{2}}^{(d-p-2)}) \delta_{d,\t{even}}\log(\frac{\ell}{\varepsilon}) \\
		    & \phantom{=~} - \frac{1}{2}\qty(\b_p(\pd\aent)+\b_{d-p-1}(\pd\aent))\log\abs{\det\bbK},
	\end{aligned}
\end{equation}
where $\mathrm{C}_k\coloneqq \mathrm{I}_k \ex{2k-d}$ and we traded the energy scale for a length scale $\ell\equiv \mu^{-1}\ex{}$ as is more common in the presentation of entanglement entropies. In the above formula, we have ignored terms with \(k>\left\lfloor\frac{d-1}{2}\right\rfloor\) (which are present only when \(d\) is odd), since they give vanishing contributions in the limit \(\varepsilon\to 0\).

\subsubsection*{On extracting universal features:}

Some comments are in order regarding our main result, \Cref{eq:EE-replica-full}. In odd dimensions we have a universal subleading correction given by
\begin{align}
	\tee= - \frac{1}{2}\qty(\b_p(\pd\aent)+\b_{d-p-1}(\pd\aent))\log\abs{\det\bbK}.
\end{align}
In even dimensions, however, the log term
\begin{align}
	\frac{\kappa}{2} \qty(\cI_{\frac{d-2}{2}}^{(p-1)}+\cI_{\frac{d-2}{2}}^{(d-p-2)})\log(\frac{\ell}{\varepsilon})
\end{align}
spoils the universality of the $\log\abs{\det\bbK}$ as rescalings of the cutoff result in constant shifts of $\ee$. Under general arguments, there is nothing that prohibits $\log$ contributions to the entanglement entropy in even dimensions and, indeed, here we find such a contribution. The coefficient of this $\log$ is a potentially universal piece of data, but in fact it depends on the geometry of the entangling surface through its heat kernel coefficient. On top of this, throughout the paper we had been neglectful of terms proportional to the Euler characteristic, \(\upchi(\pd\aent)\). The main reason is that, while topological, they ultimately arise (or can be absorbed into) ambiguities in the edge theory path integral measure and so are non-universal. This non-universality is reflected in the entropy: the Euler characteristic can be recast as the integral of a local quantity, through the generalized Gauss--Bonnet theorem. Nevertheless, it was explained in \cite{Grover:2011fa} that a generalization of the Kitaev--Preskill--Levin--Wen protocol \cite{kitaev2006topological,Levin:2006zz} is possible in higher dimensions. By this, it is meant that one can decompose the entangling surface, \(\pd\aent\), into pieces, \(\pd\aent_i\) of specific geometry. Then one can consider a linear combination of entanglement entropies across \(\pd\aent_i\), such that, when surgered together, the locally integrated contributions to the entropy cancel out. We expect, through such a construction, one can extract in all odd dimensions the topological entanglement entropy:
\begin{empheq}[box=\obox]{equation}
	\tee=-\frac{1}{2}\qty(\b_p(\pd\aent)+\b_{d-p-1}(\pd\aent))\log\abs{\det\bbK}~,
\end{empheq}
although we do not do so in this paper. In even dimensions the existence of the logarithm of the length scale of $\pd\aent$ makes local subtraction schemes subtle, although judicious schemes may still exist. Let us comment on this now.

As an example, consider four-dimensional BF theory and take \(p=1\) or \(p=2\). The two cases are interchangeable, as they are mapped to each other by interchanging the roles of the \(A\)  and \(B\) fields in the original action \Cref{eq:BFact}. Generally, it holds that $\cI^{(0)}_1 = \frac{1}{6}\upchi(\pd\aent)$ \cite{10.4310/jdg/1214427880}. Taking the entangling surface to be a torus, \(\pd R=\T^2\), for \(p=1\) we have $\cI^{(0)}_1 = 0$.  This is precisely the scenario for the first subtraction scheme\footnote{The second subtraction scheme of \cite{Grover:2011fa} involves taking $\aent$ to be a solid cylinder which also results in $\upchi(\pd\aent)=0$. They also propose an invalid subtraction scheme with $\aent$ topologically equivalent to a three-ball; it is easy to see from our result how this will result in a remaining log contribution.} proposed in \cite{Grover:2011fa}. One can explicitly show that the same holds for \(p=2\), as expected, by analyzing the spectrum of the transverse Laplacian acting on one-forms. Therefore, the topological entanglement entropy for the four-dimensional case can be uniquely specified as
\begin{align}
	\tee= -2\log\abs{\det\bbK}.
\end{align}
Note that in four dimensions the subleading part of the entropy is always topological (although possibly not universal), since \(\cI^{(p-1)}_{\frac{d-2}{2}}\) is given by the Euler characteristic. In contrast, in higher dimensions, this is not the case. In six dimensions, for example,
\begin{align*}
	\cI^{(0)}_2 = \frac{1}{180}\int_{\pd\aent} \dd\vol_{\pd \aent}\ \qty(10 \cR^2 - \cR_{\mu\nu\rho\sigma} \cR^{\mu\nu\rho\sigma} + 2 \cR_{\mu\nu} \cR^{\mu\nu}),
\end{align*}
which depends sensitively on the geometry of the entangling surface. Defining suitable Kitaev--Preskill--Levin--Wen-like subtraction scheme to extract the universal pieces of data then depends on finding regions for which this coefficient can vanish. We leave as an open question the possibility of doing this in generic even dimensions, however we hope that our result (which establishes concretely the coefficient of the logarithm) will serve as a useful guide.

\section{The extended Hilbert space and a current algebra}\label{sect:edgemodes}

Above we have presented a replica path integral calculation of the entanglement entropy.  We have focussed on the entropy that comes entirely from an \textquote{edge-mode theory,} which we have coined as \textquote{chiral mixed Maxwell,} in the high-temperature limit; this essentially counts a regulated Hilbert space dimension of this edge theory.  In this section, we provide a more honest accounting of the entanglement entropy, accounting for the subtleties of gauge invariance. This will provide an alternative view of the role of these edge-modes as providing an arena, the \textquote{extended Hilbert space} (EHS), by which the entanglement entropy can be precisely defined. We will see that a physical state is embedded as a maximally entangled state in this EHS and thus its entanglement entropy is naturally interpreted as a dimension of the EHS, which we will regulate. The upshot of this section is that we will reproduce the topological entanglement entropy from a more rigorous starting point, while also providing a, perhaps, intuitive view on the need for the edge theory and why it contributes to the entanglement entropy.

Along the way we will show that the EHS is completely organized by an infinite-dimensional spectrum-generating current algebra, entirely analogous to the Kac-Moody algebras appearing at the edge of three-dimensional topological phases. In general dimensions this algebra is non-conformal. Regardless, we will show that it is powerful enough to fix the entire computation of the entanglement entropy. We will also show that this algebra, remarkably, generates the entire spectrum of the edge-mode theory from \Cref{sec:replica}. The fact that the spectrum of the chiral Maxwell theory is entirely fixed by a (potentially non-conformal) current algebra echoes the results of \cite{Hofman:2018lfz,Hofman:2024oze} for 1-form Maxwell theories in four-dimensions. Here we present the generic story which we regard as a major result of this paper.

\subsection{Gauge-(in)variance, entanglement, and the extended Hilbert space}

The typical starting point of the entanglement entropy is the supposition of a Hilbert space factorization between a region, $\aent$, and its complement, $\coaent$:
\begin{equation}\label{eq:Hfactorsupp}
	\cH=\cH_\aent\otimes\cH_{\coaent}~.
\end{equation}
This supposition typically fails in quantum field theory due to an infinite number of correlated short-distance modes,\footnote{More modernly, and more precisely, it fails due to the type-III nature of the von Neumann algebra associated to a subregion; see e.g. \cite{Witten:2018zxz} and references therein.} however, one can imagine regulating this computation by a short-distance regulator, $\varepsilon$, (say by putting the system on a lattice, or utilizing a mutual information regulator \cite{Casini:2015woa}). However, even in a regulated scenario, \Cref{eq:Hfactorsupp}, fails for quantum gauge theories due to the global nature of gauge constraints that physical states must satisfy \cite{Buividovich:2008gq,Donnelly:2011hn,Donnelly:2014fua,Donnelly:2015hxa,Soni:2015yga,Casini:2015woa,Donnelly:2016auv}. It is, by now, well understood how to properly define entanglement entropy in gauge theories. One method is to utilize an algebraic definition of the von Neumann entropy applied to the reduced density matrix realized directly as a gauge-invariant operator \cite{Casini:2015woa}.  We have explored that aspect of entanglement entropy in this family of theories in a companion paper, \cite{Fliss:2023dze}. Here we focus on the alternative formulation, which goes by the name of the \textquote{extended Hilbert space} \cite{Buividovich:2008gq,Donnelly:2011hn,Donnelly:2014fua,Donnelly:2015hxa,Soni:2015yga}.

In short, while the physical Hilbert space, $\cH$, cannot be realized as a tensor product, we can embed $\cH$ (let us call the embedding $\cI$) into an EHS, $\cH_\t{ext}$, which admits a tensor product.
\begin{equation}
	\cH\overset{\cI}{\longhookrightarrow}\cH_\t{ext}\coloneqq\cH_\aent\otimes\cH_{\coaent}~.
\end{equation}
By definition, $\cH_\t{ext}$ is furnished with unphysical, gauge-variant states. These are states in either $\cH_\aent$ or $\cH_{\coaent}$ which carry the action of gauge transformations acting on the entangling surface, $\pa\aent$. That is to say, it is furnished with edge-mode states living on $\pa\aent$. Given a physical state $\ket{\psi}\in\cH$ it embeds to a state $\ket*{\widetilde{\psi}}\in\cH_\t{ext}$ which we can then tensor-factorize and reduce upon $\aent$:
\begin{equation}
	\widetilde{\rho}_\aent\coloneqq\Tr_{\cH_{\coaent}}\ketbra*{\widetilde{\psi}}{\widetilde{\psi}}
\end{equation}
and the entanglement entropy can defined in the typical way:
\begin{equation}
	\ee\coloneqq-\Tr_{\cH_\aent}\left(\widetilde{\rho}_\aent\log\widetilde\rho_\aent\right)~.
\end{equation}
In practice, this can be computed utilizing the replica trick we reviewed above. The important step in this process, however, is identifying the embedding map, $\cI$, and thus the appropriate embedded state $\ket*{\widetilde{\psi}}\in\cH_\t{ext}$. Since $\cH_\t{ext}$ is furnished (by definition) by gauge-variant states, this embedding is determined by demanding gauge-invariance by hand: i.e. if a gauge transformation labeled by $\alpha$ with support on $\pa\aent$ acts on $\cH_{\aent}$ and $\cH_{\coaent}$ with charge operators $Q_\aent[\alpha]$ and $Q_{\coaent}[\alpha]$, respectively, then we will demand
\begin{equation}\label{eq:QGC}
	\left(Q_\aent[\alpha]\otimes 1+1\otimes Q_{\coaent}[\alpha]\right)\ket*{\widetilde{\psi}}\overset{!}{=}0~.
\end{equation}
We will refer to \Cref{eq:QGC} as the \textquote{quantum gluing condition.}  Because $\alpha$ can be arbitrarily local to $\pa\aent$, this induces correlations to $\ket*{\widetilde{\psi}} \vphantom{\displaystyle\sum}$ that are local to $\pa\aent$ and maximally entangled. As a result $\widetilde\rho_\aent$ will be maximally mixed amongst an infinite number of edge-modes that are local to $\pa\aent$. Regulating this maximal mixture will result in a (divergent) area law (with possibly subleading area laws) to the entanglement entropy, but there may also exist universal corrections.  We will make these ideas concrete in this theory below. To begin, we will first identify a suitable $\cH_\aent$ with which to build the extended Hilbert space.

\subsection{A current algebra}

We will construct $\cH_\aent$ by quantizing the theory on \(X = \R\times \aent\).  Many of the details here follow the opening sections of \cite{Fliss:2023dze} for quantizing BF theory, however, now allowing for the existence of a boundary \(\pd \aent \neq \varnothing\). We refer the reader there for a more detailed accounting of the procedure and notation and point out only the necessary features special to the current situation here.  We will often use the embedding map of the boundary \(i_{\pd\aent}:{\pd\aent}\hookrightarrow \aent\), and \(i_{\pd X}:\pd X\hookrightarrow X\), as well as the embedding map of \(\aent\) into \(X\): \(i_\aent:\aent\hookrightarrow X\).

We split $A=A_0+a$ and $B=B_0+b$ with $A_0$ and $B_0$ with \textquote{a leg} along $\R$ and \(a\) and \(b\) along $\aent$. Correspondingly the action \Cref{eq:BFact} splits,
\begin{align}
	S_{\BF}[A_0+a,B_0+b]=\frac{\bbK}{2\pi}\int_X\left((-1)^{d-p}\sd b\wedge A_0+B_0\wedge\sd a+b\wedge da\right),
\end{align}
once we impose boundary conditions $i_{\pd X}^\ast\, A_0=i_{\pd X}^\ast B_0=0$. Above $\sd$ should be regarded as the exterior derivative along $\aent$.  $A_0$ and $B_0$ impose flatness of \(a\) and \(b\) on $\aent$:
\begin{equation}\label{eq:GLcons}
	\sd a=\sd b=0
\end{equation}
which we will refer to as the Gauss law constraints.  The residual gauge transformations, preserving the boundary conditions act on \(a\) and \(b\) as
\begin{align}\label{eq:transgaugevar}
	\var b = \sd{\beta},  & \qquad i_{\pd\aent}^* \dd_\R \beta = 0   \\
	\var a = \sd{\alpha}, & \qquad i_{\pd\aent}^* \dd_\R \alpha = 0.
\end{align}
Recall that higher-form gauge theories come with a tower of lower (secondary, tertiary, quaternary, etc.) gauge-invariances. These lower invariances indicate some redundancy in what arises as a global symmetries when $\aent$ has a boundary. Later on, we will impose restrictions on the gauge parameters that survive all the way to the boundary, to take care of this.

The charges associated to \Cref{eq:transgaugevar} can be built from the symplectic form 
\begin{align}\label{eq:sympba}
	\Omega_\aent=(-1)^{d-p-1}\frac{\bbK}{2\pi}\int_{\aent}\var b\wedge \var a.
\end{align}
Above, we view $\var b$ and $\var a$ as variational one-forms.  We define {\it variational vectors}, $\vec{v}_\alpha$ and $\vec{v}_\beta$ such that their interior product gives \Cref{eq:transgaugevar}:
\begin{align}
	\inp{\vec{v}_\alpha}{\var a}=\sd\alpha \qquad\qq{and}\qquad \inp{\vec{v}_\beta}{\var b}=\sd\beta.
\end{align}
The charges, given by
\begin{align}
	\var Q[\alpha]\coloneqq\inp{\vec{v}_\alpha}{\Omega_\aent} \qquad\qq{and}\qquad \var Q[\beta]\coloneqq \inp{\vec{v}_\beta}{\Omega_\aent},
\end{align}
can then be found as
\begin{align} \label{eq:charges}
	\begin{split}
		Q[\alpha] & \coloneqq (-1)^{d-p} \frac{\bbK}{2\pi}\int_{\aent} b\w\sd \alpha=-\frac{\bbK}{2\pi}\int_{\pd\aent}b\w\alpha\\
		Q[\beta] & \coloneqq (-1)^{d-p-1}\frac{\bbK}{2\pi}\int_\aent \sd\beta\wedge a=(-1)^{d-p-1}\frac{\bbK}{2\pi}\int_{\pd\aent}\beta\wedge a
	\end{split}
\end{align}
up to a total variation.\footnote{This total variation ambiguity doesn't affect the algebra of the charges, of course.}  The second equality in each line emphasizes that the charges localize to ${\pd\aent}$ upon imposing the Gauss law constraint, \Cref{eq:GLcons}. When these charges do not vanish identically, then they are genuine, global, symmetries of the system, i.e. they act on and transform states.  As alluded to above, there is still a need to fix an additional gauge redundancy: if $i_{\pd\aent}^\ast\alpha=\sd\gamma$ for some $\gamma\in\Omega^{p-2}({\pd\aent})$ then $Q[\alpha]$ is identically zero by pulling the Gauss constraint on \(b\) back to ${\pd\aent}$ (a similar argument follows for $\beta$).  We fix this by imposing
\begin{align}
	\dd^\dagger i_{\pd\aent}^\ast\alpha=\dd^\dagger i_{\pd\aent}^\ast\beta=0
\end{align}
This constraint fixes completely all the lower-invariances. The algebra of the charges is also given by the symplectic form as
\begin{align}
	\acomm{Q[\alpha]}{Q[\beta]}_\aent & =
	\Omega_\aent(\vec{v}_\beta,\vec{v}_\alpha)=(-1)^{d-p-1}\frac{\bbK}{2\pi}\int_\aent\sd\beta\wedge\sd\alpha
\end{align}
The canonical quantization of the charges, \(\acomm{\red\blob}{\blue\blob}\to-\ii\comm{\red\blob}{\blue\blob}\), then yields a centrally extended algebra
\begin{empheq}[box=\obox]{align}
	\comm{\hat{Q}[\alpha]}{\hat{Q}[\beta]}_\aent  &= \ii(-1)^{d-p-1}\frac{\bbK}{2\pi}\int_\aent \sd{\beta}\w\sd{\alpha} = \ii\frac{\bbK}{2\pi}\int_{\pd\aent} \sd{\beta}\w\alpha. \label{eq:KM1}
\end{empheq}

\subsection{Mode expansion of the current algebra}\label{ssec:modeexp}

We want to turn the algebra \Cref{eq:KM1} into a countable current algebra. For that we want to decompose our various forms into modes of a common basis and find their mode algebra.

In the following, we restrict ourselves within \({\pd\aent}\) and drop the bold differential. It is understood that everything below has been pulled back to \({\pd\aent}\). We will keep in mind our restriction that the forms \(\alpha\) and \(\beta\) are coclosed (though not necessarily co-exact) in ${\pd\aent}$. The set of eigen-\(k\)-forms of the transversal Laplacian, \(\eval{\cdd\dd}_{\Omega^k}\), provides a basis for the coclosed \(k\)-forms. Poincaré duality and the Hodge decomposition tell us that the non-zero spectrum of $\cdd\dd$ acting on \(k\)-forms on a \(D\)-dimensional compact manifold is the same as its spectrum acting on $(D-k-1)$-forms, since they are both related to the spectrum of $\dd\cdd$ on $(D-k)$-forms.

So, as far as the non-zero-modes are of concern, in our case, we need the bases provided by the \((p-1)\)-forms and \((d-p-2)\)-forms. Since the non-zero spectra of \(\cdd\dd\) on the \((d-2)\)-dimensional manifold \({\pd\aent}\) on these two spaces coincide, these forms will be labeled by the same index set \(\set{\sfn\in\sN_\perp^*}\). Namely, we have two orthonormal bases
\begin{align}
	\set{\varphi_\sfn\in \Omega^{p-1}({\pd\aent})}_{\sfn\in\sN_\perp^*} & \qq{and} \set{\chi_\sfn\in \Omega^{d-p-2}({\pd\aent})}_{\sfn\in\sN_\perp^*}, \\
	\ip{\varphi_\sfn}{\varphi_\sfm}                                     & = \delta_{\sfn\sfm} = \ip{\chi_\sfn}{\chi_\sfm}, \nonumber
\end{align}
with
\begin{align}
	\mqty{\cdd\dd \varphi_\sfn = \lambda_\sfn \varphi_\sfn \\[1em] \cdd\dd \chi_\sfn = \lambda_\sfn \chi_\sfn}\ , \qquad \lambda_\sfn\neq 0\ \fall\sfn\in\sN_\perp^*. \label{eq:eigeneq}
\end{align}
Now, coming to the zero-modes, these are simply the harmonic \((p-1)\)- and \((d-p-2)\)-forms on \(\pd\aent\). The natural bases to expand those are the topological bases
\begin{align}
	\set{\tau^{(p-1)}_\sfi}_{\sfi=1}^{\b_{p-1}(\pd\aent)} & \qq{and} \set{\tau^{(d-p-2)}_\sfi}_{\sfi=1}^{\b_{d-p-2}(\pd\aent)},
	\intertext{satisfying}
	\int_{\eta^\sfj} \tau^{(p-1)}_\sfi          & = \delta^\sfj_\sfi = \int_{\sigma^\sfj} \tau^{(d-p-2)}_\sfi,
\end{align}
for all \((p-1)\)- and \((d-p-2)\)-cycles \(\eta^\sfj\) and \(\sigma^\sfj\) respectively.

We can then expand:
\begin{align}
	\alpha & = \sum_{\sfn\in\sN^*_\perp} \alpha^\sfn\varphi_\sfn +\sum_{\sfi=1}^{\b_{p-1}} \alpha^\sfi \tau^{(p-1)}_\sfi   \\
	\beta  & = \sum_{\sfn\in\sN^*_\perp} \beta^\sfn \chi_\sfn + \sum_{\sfi=1}^{\b_{d-p-2}} \beta^\sfi \tau^{(d-p-2)}_\sfj.
\end{align}
Similarly, we can expand \(\hat{a}\) and \(\hat{b}\) as
\begin{align}\label{eq:ab-modeexp}
	\begin{split}
		\hat{a} &= (-1)^{d-p-1} 2\pi \qty[\inv{\bbK}]\qty(\sum_{\sfn\in\sN_\perp^*} \hat{a}_\sfn \star \chi_\sfn + \sum_{\sfi=1}^{\b_p} \hat{a}_{0\sfi} \star \tau^{(d-p-2)}_\sfi) \\
		\hat{b} &= - 2\pi \qty[\inv{\bbK}]\qty(\sum_{\sfn\in\sN_\perp^*} \hat{b}_\sfn \star \varphi_\sfn + \sum_{\sfi=1}^{\b_{d-p-1}} \hat{b}_{0\sfi} \star \tau^{(p-1)}_\sfi).
	\end{split}
\end{align}
so that the charges become
\begin{align}
	\hat{Q}[\alpha] & = \sum_{\sfn\in\sN_\perp^*} \hat{b}_\sfn \alpha^\sfn + \sum_{\sfi,\sfj=1}^{\b_{p-1}} \hat{b}_{0\sfi} \qty[\bbG_{p-1}^{\pd\aent}]_{\sfi\sfj} \alpha^\sfi    \\
	\hat{Q}[\beta]  & = \sum_{\sfn\in\sN_\perp^*} \hat{a}_\sfn \beta^\sfn + \sum_{\sfi,\sfj=1}^{\b_{d-p-2}} \hat{a}_{0\sfi} \qty[\bbG_{d-p-2}^{\pd\aent}]_{\sfi\sfj} \beta^\sfi,
\end{align}
where we note again that by Poincaré duality it holds that \(\bbG_{k}^{\pd\aent}=\qty[\bbG_{d-2-k}^{\pd\aent}]^{-1}\), up to an invertible matrix which can be set to unity by a choice of basis.

Equivalently, we can invert \Cref{eq:ab-modeexp} to write the individual modes as
\begin{align}
	\hat{a}_\sfn    & = (-1)^{d-p-1} \frac{\bbK}{2\pi} \int_{\pd \aent} \chi_\sfn \w\hat{a}                                                                                      \\
	\hat{a}_{0\sfi} & = (-1)^{d-p-1} \frac{\bbK}{2\pi} \sum_{\sfj=1}^{\b_{p}}\qty[\bbG_p^{\pd\aent}]_{\sfi\sfj}\int_{\pd \aent} \tau^{(d-p-2)}_{\sfj} \w\hat{a}                  \\[0.5em]
	\hat{b}_\sfn    & = - \frac{\bbK}{2\pi} \int_{\pd \aent} \varphi_\sfn \hat{b}                                                                                        \\
	\hat{b}_{0\sfi} & = - \frac{\bbK}{2\pi} \sum_{\sfj=1}^{\b_{d-p-1}^{\pd\aent}}\qty[\bbG_{d-p-1}^{\pd\aent}]_{\sfi\sfj}\int_{\pd \aent} \tau^{(p-1)}_{\sfj} \w\hat{b}.
\end{align}

With this mode expansion, the current algebra \Cref{eq:KM1} is written in modes as
\begin{align}
	\comm{\hat{b}_\sfn}{\hat{a}_\sfm} & = \frac{\ii\,\bbK}{2\pi}\; C_{\sfn\sfm}, \qquad \sfn,\sfm\in\sN_\perp^*, \qq{where} \label{eq:KMpre2} \\
	C_{\sfn\sfm}                      & \coloneqq \int_{\pd\aent} \dd{\chi_\sfm}\w\varphi_\sfn.
\end{align}
Of course, the zero-modes have trivial commutators with all the other modes as well as amongst themselves. The final step to pin down the algebra completely is to analyze the matrix \(C_{\sfn\sfm}\). Observe that the two bases of the non-zero sectors are related by
\begin{align}
	\dd{\chi_\sfn} = \sum_{\sfm\in\sN_\perp^*} C_{\sfm\sfn} \star\varphi_\sfm. \label{eq:connecting-bases}
\end{align}
Let us momentarily assume that there is no degeneracy of eigenvalues, i.e. \(\lambda_\sfn=\lambda_\sfm \Rightarrow \sfn=\sfm\) and act on \Cref{eq:connecting-bases} with \(\dd\cdd\). We get that for all \(\sfn\in\sN_\perp^*\) we must have
\begin{align}
	\sum_{\sfm\in\sN_\perp^*} C_{\sfm\sfn}(\lambda_\sfn-\lambda_\sfm) = 0.
\end{align}
This can only be achieved if \(C_{\sfm\sfn}\) is diagonal: \(C_{\sfm\sfn} = C_\sfm \delta_{\sfm\sfn}\). Now consider the inner-product \(\ip{\dd{\chi_\sfn}}{\dd{\chi_{\sfm}}}\). We have
\begin{align}
	 & \ip{\dd{\chi_\sfn}}{\dd{\chi_{\sfm}}} = \ip{\chi_\sfn}{\adj{\dd}\dd \chi_{\sfm}} = \lambda_{\sfm} \delta_{\sfn \sfm} \nn
	 & \phantom{---,}\roteqq                                                                                                    \\
	 & C_\sfn C_\sfm \ip{\star\varphi_\sfn}{\star\varphi_{\sfm}} = C_{\sfn}^2 \delta_{\sfn\sfm}. \nonumber
\end{align}
This tells us that \(\abs{C_\sfn} = \sqrt{\lambda_\sfn}\). We can arbitrarily choose the sign of the square root. Different signs will correspond to different choices of raising and lowering operators. We choose the following convention, which will make the choice of raising and lowering operators consistent across all \(d\) and \(p\): \(C_\sfn = \sqrt{\lambda_\sfn}\).

Returning to the general case and allowing for degeneracy of the eigenvalues, we note that the two bases only mix elements of the same eigenvalue, giving thus \(C_{\sfm\sfn}\) in a block-diagonal form. Then we can reshuffle the basis within a subspace of a fixed eigenvalue to fully diagonalize \(C_{\sfm\sfn}\) and hence the result is unchanged. Altogether we can write the final expression for the current algebra:
\begin{empheq}[box=\obox]{align}
	\comm{\hat{b}_\sfn}{\hat{a}_\sfm} & = \frac{\ii\,\bbK}{2\pi}\ \sqrt{\lambda_\sfn} \delta_{\sfn\sfm}. \label{eq:KM2}
\end{empheq}
We note the similarity to the \km\ algebras appearing in 2d, however it is important to emphasize that the modes here are not labeled, necessarily, by integers, but instead by the countable set of eigenfuctions $\sN_\perp^\star$.

\subsection{Verma modules, characters, and extended characters}

The current algebra, \Cref{eq:KM2}, allows us to build $\cH_\aent$ as a direct sum of Verma modules corresponding to integrable representations of this algebra. It will be useful to supplement this algebra with a positive-definite Hamiltonian.  Recall that since the bulk theory is first order in derivatives, its Hamiltonian is identically zero and so this is an ingredient we will add in by hand. The Hamiltonian we will supplement will be the natural generalization of the \textquote{Sugawara Hamiltonian} in 2d \km\ algebras.

Let us define ladder operators of the non-zero-modes as
\begin{equation}\label{eq:JK-currents}
	\hat{K}_\sfn \coloneqq \frac{1}{\sqrt{2}}\qty(\mu^{\Delta/2}\hat{a}_\sfn + \ii \mu^{-\Delta/2}\hat{b}_\sfn) \qq{and} \hat{J}_\sfn \coloneqq \frac{1}{\sqrt{2}}\qty(\mu^{\Delta/2}\hat{a}_\sfn - \ii \mu^{-\Delta/2}\hat{b}_\sfn)~.
\end{equation}
where $\mu$ is an energy scale and $\Delta=d-1-2p$. We then write
\begin{equation}\label{eq:Hsug}
	\hat H=\hat H_\text{zero}+\hat H_\text{osc}
\end{equation}
with
\begin{align}
	\hat{H}_\text{zero} & = \mu^\Delta \pi \inv{\bbK} \sum_{\sfi,\sfj=0}^{\b_p} \hat{a}_{0\sfi} \qty[\bbG_p^{\pd\aent}]^{-1}_{\sfi\sfj} \hat{a}_{0\sfj} + \mu^{-\Delta} \pi \inv{\bbK} \sum_{\sfi,\sfj=0}^{\b_{d-p-1}} \hat{b}_{0\sfi} \qty[\bbG_{d-p-1}^{\pd\aent}]^{-1}_{\sfi\sfj} \hat{b}_{0\sfj}\nonumber \\
	\hat{H}_\t{osc}     & = \pi\inv{\bbK} \sum_{\sfn\in\sN_\perp^*}\qty(\mu^\Delta \hat{a}_\sfn \hat{a}_\sfn + \mu^{-\Delta} \hat{b}_\sfn \hat{b}_\sfn) =  2\pi\inv{\bbK} \sum_{\sfn\in\sN_\perp^*} \hat{J}_\sfn \hat{K}_\sfn + \mu E_0,
\end{align}
with $E_0$ a potentially divergent zero-point energy that will play no role in the entanglement entropy. This Hamiltonian is motivated by three very important points: firstly, it is positive-definite, and secondly, when expressed in terms of the ladder-operators it plays a natural algebraic role, extending \Cref{eq:KM2} to
\begin{align}
	\comm{\hat{J}_\sfn}{\hat{K}_\sfm}                               & = \frac{\bbK}{2\pi}\sqrt{\lambda_\sfn} \delta_{\sfn\sfm} \\
	\comm{\hat{H}}{\hat{K}_\sfn} = \sqrt{\lambda_\sfn} \hat{K}_\sfn & \qq{and}
	\comm{\hat{H}}{\hat{J}_\sfn} = -\sqrt{\lambda_\sfn} \hat{J}_\sfn. \label{eq:ladder-comms}
\end{align}
It is evident in this writing, that the operator \(\hat{J}_\sfn\), with \(\sfn\in\sN_\perp^*\), lowers the energy by \(\sqrt{\lambda_\sfn}\) units and \(\hat{K}_\sfn\) raises the energy by the same amount.

Thirdly, and perhaps most importantly, when expressed in terms of the original field variables on $\pa\aent$, $\hat H$ is \emph{local}. Namely writing
\begin{align}
	i_{\pd X}^*A & = a+\dd{t}\w \widetilde{a}, \\
	i_{\pd X}^*B & = b+\dd{t}\w \widetilde{b},
\end{align}
with \(a\in \Omega^0(\R)\otimes\Omega^p(\pd \aent)\), \(\widetilde{a}\in \Omega^0(\R)\otimes\Omega^{p-1}(\pd \aent)\), \(b\in\Omega^0(\R)\otimes\Omega^{d-p-1}(\pd \aent)\), and \\ \(\widetilde{b}\in\Omega^0(\R)\otimes\Omega^{d-p-2}(\pd \aent)\), $\hat H$ takes the form
\begin{align}\label{eq:Hlocal}
	\hat{H} = \frac{\mu^\Delta}{4\pi}\ip{a}{\bbK\,a}_{\pd \aent} + \frac{\mu^{-\Delta}}{4\pi} \ip{b}{\bbK\,b}_{\pd \aent}.
\end{align}
One can already recognize $\hat H$ as the Hamiltonian of the chiral mixed Maxwell theory appearing in \Cref{sec:replica}.\footnote{Indeed, this matching is made explicit by \textquote{adding in} the time component, $\widetilde{a}$, of $i_{\pd X}^* A$ to $\hat H$ by replacing \(b\) through the \textquote{chiral} boundary condition $b = \mu^\Delta (-1)^{(d-p)p}\star_{\scriptscriptstyle\pd \aent}\widetilde{a}$.}

The structure of the $\cH_\aent$ factor of the extended Hilbert space is now clear. The eigenstates of the zero-modes \(\hat{a}_{0\sfi}\) and \(\hat{b}_{0\sfj}\) define primary, lowest-weight states, \(\ket{\omega}\), that are annihilated by all \(\hat{J}_\sfn\). We can act on each such state with \(\hat{K}_\sfn\), repeatedly to construct a whole Verma module,
\begin{align}\label{eq:verma}
	\cV_\omega & \coloneqq \operatorname{span}\set{\ket{\omega,\set{N_\sfn}}} \coloneqq \operatorname{span}\set{\prod_{\sfn\in\sN_\perp^*}\prod_{\sfI=1}^\kappa \qty(\hat{K}_{\sfI\sfn})^{N_{\sfI \sfn}}\ket{\omega}}.
\end{align}
The full Hilbert space is then a direct sum of all Verma modules
\begin{align}\label{eq:HSasVerma}
	\cH_{\aent} = \bigoplus_\omega \cV_\omega.
\end{align}
The primary states are labeled by the eigenvalues of the current algebra zero-modes. These we can easily find by using Dirac quantization restricting to \(\pd \aent\):
\begin{align}\label{eq:flux-quant-a}
	\int_{\sigma_\sfi} a = 2\pi n^{\sfi},\qquad \fall \sigma_\sfi\in\H_p(\pd\aent) \qquad n^\sfi\in\Z~,
\end{align}
and
\begin{align}\label{eq:flux-quant-b}
	\int_{\eta_\sfi} b = 2\pi m^{\sfi},\qquad \fall \eta_\sfi\in\H_p(\pd\aent) \qquad m^\sfi\in\Z.
\end{align}

Now, invoking the mode expansion, \Cref{eq:ab-modeexp}, this amounts to
\begin{align}
	\hat{a}_{0\sfi}\ket{\omega} & = \bbK\sum_{\sfj=1}^{\b_p} n^\sfj \qty[\bbG_p^{\pd\aent}]_{\sfi\sfj}\ket{\omega}                            \\
	\hat{b}_{0\sfi}\ket{\omega} & = \bbK\sum_{\sfj=1}^{\b_{d-p-1}} m^\sfj \qty[\bbG_{d-p-1}^{\pd\aent}]_{\sfi\sfj}\ket{\omega}.
\end{align}
Therefore, the primary states, are labeled by two vectors of integers, \(\vec{n}\in\Z^{\kappa\,\b_{p}(\pd\aent)}\) and \(\vec{m}\in\Z^{\kappa\,\b_{d-p-1}(\pd\aent)}\), \(\ket{\omega}\equiv\ket{\vec{n},\vec{m}}\), and have energy
\begin{align}\label{eq:primaryenergy}
	\Delta_{\vec{n}\vec{m}} = \pi \mu\qty(\vec{n}\cdot\qty(\bbK\otimes\widetilde{\bbG}_p^{\pd\aent})\cdot\vec{n} + \vec{m}\cdot\qty(\bbK\otimes\widetilde{\bbG}_{d-p-1}^{\pd\aent})\cdot\vec{m}),
\end{align}
where we have, again absorbed the powers of $\mu$ into the dimensionless Gram matrices $\widetilde{\bbG}_k^{\pd\aent}$. From here on we will also measure our energies in units of $\mu$, i.e. $\sqrt{\lambda_{\sf n}}\rightarrow\mu\sqrt{\lambda_{\sf n}}$ with the new $\lambda_{\sfn}$'s dimensionless.

We can now define extended characters of our algebra. Having decomposed the Hilbert space as
\begin{align}
	\cH_{\aent} = \bigoplus_{\substack{\vec{n}\in\Z^{\kappa\,\b_{p}(\pd\aent)} \\ \vec{m}\in\Z^{\kappa\,\b_{d-p-1}(\pd\aent)}}} \cV_{\vec{n},\vec{m}},
\end{align}
we can first compute the characters of each Verma module, \(\ch_{\cV_{\vec{n},\vec{m}}}[q] \coloneqq \tr_{\cV_{\vec{n},\vec{m}}} q^{\hat H/\mu}\). This is straightforward. Remembering that \(\hat{K}_\sfn\) raises the energy by \(\mu\sqrt{\lambda_\sfn}\) we obtain
\begin{align}\label{eq:character}
	\ch_{\cV_{\vec{n},\vec{m}}}[q] \coloneqq \textcolor{blue}{\underset{\smqty{\t{vacuum} \\ \t{contribution}}}{\underbrace{\phantom{-} \vphantom{\qty(\prod_{\sfn\in\sN_\perp^*})}\textcolor{black}{q^{\kappa E_0}}\phantom{-}}}}\ \textcolor{red}{\underset{\text{descendants}}{\underbrace{\textcolor{black}{\qty(\prod_{\sfn\in\sN_\perp^*} \sum_{N_\sfn = 0}^\infty q^{N_\sfn \sqrt{\lambda_\sfn}})^\kappa}}}}\ \textcolor{green}{\underset{\text{primaries}}{\underbrace{\phantom{-}\vphantom{\qty(\prod_{\sfn\in\sN_\perp^*})}\textcolor{black}{q^{\Delta_{\vec{n},\vec{m}}}}\phantom{-}}}},
\end{align}
Finally, summing over all the Verma modules gives us an {\it extended character}:
\begin{align}\label{eq:extchar}
	\vec{\ch_0}[q] & \coloneqq\sum_{\substack{\vec{n}\in\Z^{\b_p}                                                                                                       \\ \vec{m}\in\Z^{\b_{d-p-1}}}}\ch_{\cV_{\vec{n},\vec{m}}}[q] \nn
	               & = q^{\kappa E_0}\qty(\prod_{\sfn\in\sN_\perp^*} \sum_{N_\sfn = 0}^\infty q^{N_\sfn \sqrt{\lambda_\sfn}})^\kappa\sum_{\substack{\vec{n}\in\Z^{\b_p} \\ \vec{m}\in\Z^{\b_{d-p-1}}}}q^{\Delta_{\vec{n},\vec{m}}}.
\end{align}
Here, we once again encounter the generalized Dedekind eta function from \Cref{sec:replica};
\begin{align}
	\upeta^{(p-1)}_{\pd\aent}[q] \coloneqq q^{-\half E_0}\qty(\prod_{\sfn\in\sN_\perp^*} \sum_{N_\sfn = 0}^\infty q^{N_\sfn \sqrt{\lambda_\sfn}})^{-1/2}~.
\end{align}
Additionally, recalling the form of the primary state energies, \Cref{eq:primaryenergy}, the sum over the primaries of $q^{\Delta_{\vec n,\vec m}}$ organizes into the Siegel Theta functions appearing in \Cref{sec:replica}.  Consequently, we find a curious and powerful correspondence: the partition function, \Cref{eq:parti-edge-1}, of the chiral mixed Maxwell theory, from \Cref{sec:replica} takes the form of an extended character of our current algebra!
\begin{empheq}[box=\obox]{align}\label{eq:edgecharcorr}
	\vec{\ch_0}[q] =\frac{\Theta\qty[q;\bbK\otimes\widetilde{\bbG}_p^{\pd\aent}]}{\qty(\upeta^{(p-1)}_{\pd\aent}[q])^\kappa}\ \frac{\Theta\qty[q;\bbK\otimes\widetilde{\bbG}_{d-p-1}^{\pd\aent}]}{\qty(\upeta^{(d-p-2)}_{\pd\aent}[q])^\kappa}=\parti_{\edge}\qty[\S^1_\varepsilon\times {\pd\aent}].
\end{empheq}

\subsection{The entanglement entropy}
We have now finally set the stage to calculate the entanglement entropy. Consider partitioning the Cauchy slice, \(\Sigma\), into two parts \(\Sigma = \aent \cup_{\pd\aent} \co{\aent}\), where \(\cup_{\pd\aent}\) denotes gluing along their common boundary. We would like to compute the entanglement entropy of a state in \(\cH_\Sigma\) upon tracing out the degrees of freedom on \(\co{\aent}\). However, we have seen that \(\dim\cH_{\Sigma}=\abs{\det\bbK}^{\b_p(\Sigma)}\), which is obviously finite, but each of \(\cH_\aent\) and \(\cH_{\co{\aent}}\) is a sum of Verma modules, \Cref{eq:HSasVerma}, and thus infinite-dimensional. Thus we starkly see that \(\cH_\Sigma\neq\cH_\aent\otimes \cH_{\co{\aent}}\). Instead, we will embed $\cH_\Sigma$ inside the tensor product and impose the quantum gluing condition, \Cref{eq:QGC}, to ensure that gauge transformations acting on $\pa R$ annihilate the physical states living in $\cH_\Sigma\subset\cH_\aent\otimes\cH_{\co\aent}$. Let us denote
\begin{align}\
	\hat{Q}_\aent[\alpha] & \coloneqq (-1)^{d(p+1)}\frac{\bbK}{4\pi} \int_{\pd \aent} \alpha \w \hat{b} = \sum_{\sfn\in\sN_\perp^*} \alpha^\sfn \hat{b}_\sfn \qq{and} \\
	\hat{Q}_\aent[\beta]  & \coloneqq \frac{\bbK}{4\pi} \int_{\pd \aent} \beta \w \hat{a} = \sum_{\sfn\in\sN_\perp^*} \beta^\sfn \hat{a}_\sfn,
\end{align}
the operators with support on \(\pd \aent\).  Similarly for $\co\aent$ there are charge generators
\begin{align}
	\hat{Q}_{\co{\aent}}[\overline{\alpha}]    & \coloneqq (-1)^{d(p+1)}\frac{\bbK}{4\pi} \int_{\pd \co{\aent}} \overline{\alpha} \w \hat{b} = \sum_{\sfn\in\sN_\perp^*} \overline{\alpha}^\sfn \hat{b}_\sfn \qq{and} \\
	\hat{Q}_{\co{\aent}}\qty[\overline{\beta}] & \coloneqq \frac{\bbK}{4\pi} \int_{\pd \co{\aent}} \overline{\beta} \w \hat{a} = \sum_{\sfn\in\sN_\perp^*} \overline{\beta}^\sfn \hat{a}_\sfn,
\end{align}
with support on \(\pd \co{\aent}\). Upon gluing \(\aent\) to \(\co{\aent}\) along their common boundary, we will identify \(\alpha\) and \({\beta}\) with their parity opposites, \(\overline{\alpha}\) and \(\overline{\beta}\), respectively. We can then define the generators of gauge transformations on the entire \(\Sigma\) as
\begin{align}
	\hat{\cQ}[\alpha] & \coloneqq \hat{Q}_\aent[\alpha]\otimes \hat{1}_{\co{\aent}} +  \hat{1}_{\aent}\otimes\hat{Q}_{\co{\aent}}[\overline{\alpha}], \\
	\hat{\cQ}[\beta]  & \coloneqq \hat{Q}_\aent[\beta]\otimes \hat{1}_{\co{\aent}} +  \hat{1}_{\aent}\otimes\hat{Q}_{\co{\aent}}\qty[\overline{\beta}],
\end{align}
where \(\hat{1}_\aent\) \(\qty(\hat{1}_{\co{\aent}})\) is the identity operator on \(\cH_\aent\) \(\qty(\cH_{\co{\aent}})\). The condition that the physical states should be uncharged under the charge generators, identifies \(\cH_\Sigma\) as a specific subspace of \(\cH_\aent\otimes\cH_{\co{\aent}}\), namely
\begin{align}
	\cH_\Sigma = \ker\hat{\cQ}[\alpha]\cap\ker\hat{\cQ}[\beta] \subset \cH_\aent\otimes\cH_{\co{\aent}}.
\end{align}
At the level of the ladder operators, \Cref{eq:JK-currents}, the quantum gluing condition can be expressed as
\begin{align}\label{eq:currentQGC}
	\qty(\hat{J}_\sfn \otimes\hat{1}_{\co{\aent}}+\hat{1}_\aent\otimes\hat{K}_\sfn)\ket{\psi} & \demeqq 0  \\
	\qty(\hat{K}_\sfn \otimes\hat{1}_{\co{\aent}}+\hat{1}_\aent\otimes\hat{J}_\sfn)\ket{\psi} & \demeqq 0.
\end{align}
which roughly states that raising operators on $\aent$ are lowering operators on $\coaent$ and vice versa.  Due to this, it is clear that $\ket*{\widetilde{\psi}}$ will be {\it maximally entangled} in terms of the oscillator occupation numbers, $\set{N_\sfn}$.  The quantum gluing condition, \Cref{eq:QGC}, applied at the level of zero-modes enforces a matching of the integer charges.  The result of this is that physical states of this $\cH_\Sigma$ embed into {\it generalized Ishibashi states} of $\cH_\aent\otimes\cH_{\co\aent}$:
\begin{align}
	\ket*{\widetilde{\psi}}=\kket{\vec 0} \coloneqq \sum_{\vec n,\vec m}\sum_{\set{N_\sfn}} \ket{\vec{n}, \vec{m}, \set{N_\sfn}}_\aent\otimes\overline{\ket{-\vec{n}, -\vec{m}, \set{N_\sfn}}_{\co{\aent}}},
\end{align}
where the overline denotes an anti-linear conjugate of the state.\footnote{We have written this for \(\kket{\vec{0}}\), corresponding to the ground state without any Wilson surfaces piercing \(\Sigma\). The computation with a generic \(\kket{\vec{r}}\) can be easily obtained, by shifting the argument of the Siegel Theta functions by a constant fractional charge.  This will not affect the end entanglement entropy.} The reduced density matrix is given by tracing over the $\cH_{\coaent}$ tensor factor:
\begin{align}
	\widetilde\rho_\aent \coloneqq \tr_{\cH_{\co{\aent}}} \kket{\vec{0}}\!\bbra{\vec{0}}. \label{eq:mixed-rho}
\end{align}
However the above object is not well defined, per se.  We need to first address the subtle issue of normalizability.  Given that $\kket{\vec{0}}$ is maximally entangled over an infinite number of modes (labeled by occupation numbers, $N_\sfn$), its norm is divergent.  To obtain a normalizable vector in $\cH_\aent\otimes\cH_{\coaent}$ we need to first regularize $\kket{\vec{0}}$ and we will do so with the Sugawara-type Hamiltonian we discussed earlier, \Cref{eq:Hsug}:
\begin{align}\label{eq:regIshi}
	\kket{\varepsilon}\coloneqq \exp(-\frac{1}{4} \varepsilon \left(\hat H_\aent+\hat H_{\coaent}\right))\kket{\vec{0}},
\end{align}
where we have explicitly denoted which factor of $\cH_\text{ext}$ the Hamiltonians act.

From here it is easy to see that the regulated Ishibashi state, \Cref{eq:regIshi}, takes the form of a canonical purification of a thermal density matrix with inverse temperature $\varepsilon$. This thermal density matrix is exactly the (unnormalized) reduced density matrix
\begin{align}\label{eq:regredDM}
	\widetilde\rho_\aent(\varepsilon) \coloneqq \tr_{\cH_{\co{\aent}}} \ex{-\frac{\varepsilon}{4}\left(\hat H_\aent+\hat H_{\coaent}\right)}\kket{\varepsilon}\!\bbra{\varepsilon}\ex{-\frac{\varepsilon}{4}\left(\hat H_\aent+\hat H_{\coaent}\right)}
\end{align}
and whose norm is given by the extended character, \Cref{eq:extchar}
\begin{align}
	\tr_{\cH_{\aent}}\widetilde\rho_\aent(\varepsilon)=\vec{\ch_0}[q_\varepsilon],\qquad\qquad q_\varepsilon\coloneqq \ex{-\varepsilon\mu}. \label{eq:norm-reduced-dens}
\end{align}
As $\varepsilon\rightarrow 0$ this is simply the regulated dimension of $\cH_\aent$. Thus, the von Neumann entropy of the reduced density matrix is the regulated $\log\dim\cH_\aent$.  More suggestively, however, given the correspondence \Cref{eq:edgecharcorr}, this is {\it also} the high-temperature limit of the thermodynamic entropy of the edge-mode theory, i.e.
\begin{align}
	\ee = \lim_{\varepsilon\to 0} \qty(1-\varepsilon\pd_\varepsilon)\parti_{\edge}\qty[\S^1_\varepsilon\times \pd\aent]. \label{eq:th-entr}
\end{align}
At this point the technical computations follow that of \Cref{sec:edgethyentanglementcalc}.  We arrive again at the main result for the entanglement entropy:
\begin{equation}\label{eq:EE-KM-prefinal}
	\begin{aligned}
		\ee & = \sum_{k=1}^{\left\lfloor\frac{d-1}{2}\right\rfloor} \mathrm{C}^{(p-1)}_k \qty(\frac{\ell}{\varepsilon})^{d-2k} + \frac{\kappa}{2} \qty(\cI_{\frac{d-2}{2}}^{(p-1)}+\cI_{\frac{d-2}{2}}^{(d-p-2)}) \delta_{d,\t{even}}\log(\frac{\ell}{\varepsilon}) \\
		    & \phantom{=~} - \frac{1}{2}\qty(\b_p+\b_{d-p-1})\log\abs{\det\bbK}.
	\end{aligned}
\end{equation}
where $\mathrm C^{(p-1)}_k$ are non-universal, dimensionless, numbers, \(\cI_{\frac{d-2}{2}}^{(k)}\) is the \(\qty(\frac{d-2}{2})\)-th heat kernel coefficient for the spectral zeta function, and we have exchanged $\ell=\mu^{-1}\ex{}$ as a characteristic length scale (see \Cref{sec:replica} for details).

\section{Discussion}\label{sec:disc}

In this paper, we considered the edge contributions to the entanglement entropy in higher-dimensional Abelian topological phases described by \(p\)-form BF theories. These are phases whose ground states are condensates of \(p\)-form surface operators. We found that the entanglement entropy coming from localized edge-modes at the entangling surface takes the form of a non-universal, divergent, area law decreasing in powers of two with a possible log divergence in even dimensions. The constant corrections to this area law are given in terms of topological features of the entangling surface, namely its $(p-1)$-st and $(d-p-2)$-st Betti numbers. Our result is upheld through two separate, but complementary, fronts: we have performed a replica path integral calculation where the entropy arises as a high-temperature thermal entropy of an edge-mode partition function living on a regulated entangling surface (with the regulator playing the role of the inverse temperature). This is edge-mode theory is a chiral combination of $(p-1)$- and $(d-p-2)$-form Maxwell theories, which we call \textquote{chiral mixed Maxwell theory.} We followed this calculation with a more rigorous definition of the entanglement entropy through an extended Hilbert space and showed that this extended Hilbert space is organized by a novel infinite-dimensional current algebra, which has not appeared in the literature (to this degree of generality) before. We elucidated features of this current algebra and extracted the entanglement entropy through its representation characters.  Along the way we have shown that these characters account for the spectrum of the Maxwell edge-theory and match its thermal partition function.

There are several features of our main result that require elaboration and that we would like to highlight at this point.

\subsubsection*{Comparing to the GTV result}

Let us comment on the discrepancy of our result with the result found by Grover, Turner, and Vishwanath \cite{Grover:2011fa} in states of discrete gauge theories described by condensates of \(p\)-form membranes:
\begin{equation}\label{eq:GTVresult}
	\ent_\text{GTV}=\ent_\text{local}-\sum_{n=0}^{p-1}(-1)^{p-1+n}\b_n(\pa\aent)\;\log{\abs{G}}
\end{equation}
where $\ent_\text{local}$ is built out of integrating local quantities (and so includes possible logs and Euler characteristics in even dimensions), and $\abs{G}$ is the order of a discrete gauge group (this is the analogue of $\abs{\det\bbK}$ in our computation). This result was arrived at by counting the constraints implied by the intersection of \(p\)-form membranes with the entangling surface.\footnote{We are grateful the authors of \cite{Grover:2011fa} for correspondence and explaining their result to us.} In order to compare \Cref{eq:mainresult1} and  \Cref{eq:GTVresult} and better highlight the discrepancy, it is useful to write our result in terms of the analogous alternating sum. We can do this either in the language of differential forms or in the language of chains.  The calculations from \Cref{sec:replica,sect:edgemodes} are natural in the language of differential forms, so let us start there.  We note
\begin{align}\label{eq:guessGTVforms}
	-\frac{1}{2}\left(\b_{p-1}+\b_{d-p-2}\right)= & -\sum_{n=0}^{p-1}(-1)^{p-1+n}\b_n\nonumber                                                                            \\
	                                              & -\frac{1}{2}\sum_{n=0}^{p-2}(-1)^{p-2+n}\dim\Omega^n-\frac{1}{2}\sum_{n=0}^{d-p-3}(-1)^{d-p-3+n}\dim\Omega^n\nonumber \\
	                                              & +\frac{1}{2}\dim \mathsf{E}^{p-1}+\frac{1}{2}\dim \mathsf{E}^{d-p-2}~,
\end{align}
where $\dim \Omega^k$ and $\dim \mathsf{E}^k$ are the dimensions of all \(k\)-forms and exact \(k\)-forms on $\pa\aent$, respectively; these are divergent quantities but can be regulated, say, on a lattice and regarding them as cochains. This equality follows from the short exact sequences.
\begin{equation}
	0\to\sfC^k\to\Omega^k(\Sigma)\xto{\dd}\mathsf{E}^{k+1}\to0~,\qquad0\to\mathsf{E}^k\to\sfC^k\to\H^k\to0~,
\end{equation}
where $\sfC^k$ is the space of closed \(k\)-forms, and utilizing Poincaré duality on $\pa\aent$, $\b_{k}=\b_{d-2-k}$.  The first line of \Cref{eq:guessGTVforms} is the desired GTV result.  It is plausible that the second line involving $\dim\Omega^n$ can be absorbed into the definition of the path integral measures for the gauge fields and the ghosts (which also take an alternating form).\footnote{See \cite{Witten:1995gf} for an inspirational manipulation of this sort in four-dimensional Maxwell theory.} However, it is harder for us to argue away the third line involving $\dim\mathsf{E}^{p-1 / d-p-2}$. This term would not be there if the object of interest were instead $-\frac{1}{2}(\dim\sfC^{p-1}+\dim\sfC^{d-p-2})$, i.e. a counting of closed forms as opposed to harmonic forms.  However, this appears unnatural in our approach.  For instance, recall that in the computation from \Cref{sec:replica}, the $\abs{\det\bbK}$ arises hand-in-hand from an instanton sum and a counting of zero-modes of the Hodge Laplacian restricted to transversal $(p-1)/(d-p-2)$-forms. Both of these objects are counted by harmonic forms, not closed forms. This is mirrored in the computation of \Cref{sect:edgemodes} where the $\abs{\det\bbK}$ arises from counting zero-modes of the current algebra. There we argued that exact forms give rise to exactly zero charges and so again a true count of the zero-modes naturally lands upon the $\b_{p-1}+\b_{d-p-2}$.

In terms of chains we can also express, through wholly similar manipulations,
\begin{align}\label{eq:guessGTVchains}
	-\frac{1}{2}\left(\b_{p-1}+\b_{d-p-2}\right)= & -\sum_{n=0}^{p-1}(-1)^{p-1+n}\b_n\nonumber                                                                          \\
	                                              & +\frac{1}{2}\sum_{n=0}^{p-1}(-1)^{p-1+n}\dim \sfC_n+\frac{1}{2}\sum_{n=0}^{d-p-2}(-1)^{d-p-2+n}\dim \sfC_n\nonumber \\
	                                              & -\frac{1}{2}\dim\mathsf Z_{p-1}-\frac{1}{2}\dim\mathsf Z_{d-p-2}~,
\end{align}
where $\dim \sfC_n$ and $\dim \mathsf Z_n$ is the dimension of all \(n\)-chains on $\pa\aent$ and the dimension of \(n\)-chains on $\pa\aent$ without boundary, respectively. Again, the first line of \Cref{eq:guessGTVchains} is commensurate with the GTV result. The second line is formed of locally integrated quantities on $\pa\aent$ and it is feasible that they can be subtracted through a Kiteav--Preskill--Levin--Wen-like scheme. It is not clear whether the third line can be locally subtracted (we indeed believe not). The GTV result would follow if instead the object of interest were $\frac{1}{2}\left(\mathsf B_{p-1}+\mathsf B_{d-p-2}\right)$ where $\mathsf B_k=\mathsf Z_k-\b_k$ is the number of boundary-less \(k\)-chains that are the boundary of a $k+1$-chain. This counting, which is in fact the one undertaken in \cite{Grover:2011fa}, is pictorially natural when viewing the ground state as a \(p\)-membrane condensate.

Further elucidating the origin of this gap is ongoing work. In this vein, having another independent mode of calculation, e.g. a higher-dimensional version of surgery, would help clarify things; see the discussion below.

\subsubsection*{On possible bulk contributions and essential topological entanglement}

In this paper we have focussed on the contribution from edge-modes to the entanglement entropy. Since this theory is topological, it is natural to believe that this is the sole contribution to the entropy; however, let us revisit this assumption. In the replica computation of \Cref{sec:replica}, we did not undertake the computation of
\begin{equation}
	\parti_{\text{bulk}}[\rep{n}]=\abs{\det\bbK}^{\h_p\qty(\rep{n},\brep{n})}\tb_\text{RS}[\rep{n},\pa\rep{n}]^{(-1)^{p-1}}
\end{equation}
but schematically, since all of the oscillator contributions to $\tb_\text{RS}$ come with Dirichlet boundary conditions, we expect them to scale with \(n\) and not contribute to the entropy. Instead, we need to worry about homological contributions of $\parti_\text{bulk}$, which can arise from cycles that can either pull back to or anchor on $\brep{n}$.

This counting is actually a bit cleaner in the extended Hilbert space discussion of \Cref{sect:edgemodes}. There we focussed on variational charges, however there are also charge operators that cannot be written variationally. These are homological surface operators
\begin{equation}
	\hW_{\eta^\sfj}^{\vec w_\sfj}\coloneqq  \exp(\int_{\eta^\sfj} w_\sfj^\sfJ\ A_\sfJ)~,\qquad \hV_{\sigma^\sfi}^{\vec v_\sfi}\coloneqq \exp(\int_{\sigma^\sfi} v_\sfi^\sfI\ B_\sfI),
\end{equation}
where $\eta^\sfj$ and $\sigma^\sfi$ are basis \(p\)- and $(d-p-1)$-homology cycles in $\aent$. These commute with the variational charges of \Cref{sect:edgemodes}, however, do not commute with each other. Thus states in the $\mathcal H_\aent$ factor of extended Hilbert space should in fact also be labeled with the eigenvalues of one set of the operators. Without loss of generality, we can label the states by $w_\sfj$ for each \(p\)-cycle in $\aent$: $\ket{\{w_\sfj\}}$. What happens to these quantum numbers when we solve for the true ground state, $\ket*{\widetilde{\psi}}$, in $\mathcal H_\aent\otimes\mathcal H_{\coaent}$? For $w_\sfj$'s corresponding to cycles that live in the interior of $\aent$ or $\coaent$, i.e. those that do not pull back to or intersect $\pa\aent$, we simply match those charges with the charge of $\ket{\psi}$ for the corresponding cycles in $\Sigma$. However, for cycles of $\aent$ that pull back to $\pa\aent$, they are unfixed: they must combine with a similar cycle from $\co\aent$ to give the correct charge corresponding to a cycle in $\Sigma$. This suggests that $\ket*{\widetilde{\psi}}$ is maximally correlated over cycles of $\Sigma$ that pull back\footnote{Cycles that intersect $\pa\aent$ transversally must additionally \textquote{match up} with a cycle from $\co\aent$ but they are not cancelled, instead their charge is fixed by the charge in $\ket{\psi}$. If there is a mismatch of either type of cycles between $\aent$ and $\coaent$ they must be set to zero by hand, i.e. if an anchored cycle of $\aent$ cannot be \textquote{completed} to a true cycle of $\Sigma$ upon gluing on $\coaent$, or in a cycle of $\aent$ that pulls back to $\pa\aent$ becomes trivial in $\co\aent$, then we simply set that charge to zero in $\ket*{\widetilde{\psi}}$.} to $\aent$. This strongly suggests that there is, in fact, a bulk entropy equal to $\abs{\det\bbK}$ times the number of such \(p\)-cycles.  This number was precisely calculated in \cite{Fliss:2023dze} and is the magnetic \(p\)-form essential topological entanglement:
\begin{align}\label{eq:Sbulkguess}
	\mathcal S_\text{bulk}\overset{?}{=} & \mathcal E_{\text{mag}}\\
	=                                    & \Bigg[\sum_{n=0}^{p-1}(-1)^{p-1-n}\b_n(\pa\aent)+\sum_{n=0}^p\left(\b_n(\Sigma)-\dim\H_n(\Sigma,\pa\aent)\right)\Bigg]\log\abs{\det\bbK}.
\end{align}
where $\H_k(\Sigma,\pa\aent)$ are relative homology classes. Interestingly the GTV alternating sum makes an appearance (albeit with the opposite sign than in \cite{Grover:2011fa}), however we do not regard this as accounting for the above discrepancy.  For instance, when $\Sigma$ is topologically trivial (say $\mathbb R^{d-1}$ or $\S^{d-1}$) then the terms in \Cref{eq:Sbulkguess} exactly cancel and $\mathcal S_\text{bulk}$ vanishes (this is by design since there are no non-trivial cycles in $\Sigma$ to \textquote{count}). Thus, the above discrepancy remains in this simple example.

\subsubsection*{Surgery}

Lastly, let us comment on the possibility of another, independent, manner of evaluating this entanglement entropy using surgery. We can proceed via replica path integral, very much in the spirit of \Cref{sec:replica}; however, instead of regulating $X_n$ by excising a tubular neighborhood, we work to directly evaluate $\parti[X_n]$ on the branched cover over $\pa\aent$. For generic manifolds this seems quite difficult (although there may be some feasible benchmark examples, e.g. when $\Sigma$ is a product of spheres and $\aent$ is a product of a disc and spheres). However, one promising avenue is to develop a program to evaluate such manifolds systematically. Let us recall the procedure in three dimensions, which hinges upon the fact that the Hilbert space on a two-sphere is one-dimensional. Thus, the path integral on any three-geometry, \(M\), with $\pd M=\S^2$, produces a state proportional to the one produced by the path integral on a three-ball, $\B^3$:
\begin{equation}
	\ket{M}\propto\ket{\B^3}~.
\end{equation}
Thus, for any manifold that can be written as union of two manifolds across a common two-sphere, $X=M_1\cup_{\S^2}M_2$, we find formally
\begin{equation}
	\parti[X]=\braket{M_1}{M_2}_{\cH_{\S^2}}=\frac{\braket{M_1}{\B^3}_{\cH_{\S^2}}\braket{\B^3}{M_2}_{\cH_{S^2}}}{\braket{\B^3}{\B^3}_{\cH_{\S^2}}}=\frac{\cZ[\overline M_1]^\ast\cZ[\overline M_2]}{\cZ[\S^3]}
\end{equation}
where $\overline M_{1,2}$ are $M_{1,2}$ with their $\S^2$ boundaries \textquote{filled-in} with a $\B^3$. This effectively allows one to \textquote{cut open} path integrals defined on complicated manifolds along two-spheres and \textquote{cap them off} smoothly and evaluate them through simpler \textquote{ingredient} path integrals, e.g. $\parti[\S^3]$. In higher dimensions, topology is more involved, but we also potentially have more tools at our disposal: equation \Cref{eq:BFHSdim} indicates that there are potentially multiple choices of $\Sigma$ with $\dim\mathcal H_\Sigma=1$ which can provide an \textquote{ingredient} for surgery. Developing this further can provide an independent check on both of the above points: the discrepancy with the GTV result as well as the existence of bulk contributions.

\section*{Acknowledgments}

It is a pleasure to thank Dionysios Anninos, Sean Hartnoll, Diego Hofman, Keivan Namjou, Onkar Parrikar, Ronak Soni, Aron Wall, and Chen Yang for discussions. We are grateful to the authors of \cite{Grover:2011fa} (particularly Tarun Grover) for correspondence. JRF thanks Rob Leigh and Matthew Lapa for early conversations that became the germ of this work. 
JRF also thanks the University of Amsterdam for hospitality. SV also thanks the University of Cambridge, 
and the ICISE in Quy Nhon, Vietnam for hospitality. JRF is supported by STFC consolidated grant ST/T000694/1 and by Simons Foundation Award number 620869. SV is supported by the NWO Spinoza prize awarded to Erik Verlinde.

\appendix

\section{Partition functions of higher-form gauge theories}\label{app:parties}

In this appendix, we provide details on the computations of the edge-mode partition function of \Cref{sec:replica}.

\subsection{The partition function of \texorpdfstring{$(p-1)$}{(p-1)}-form Maxwell theory}

First we will quantize, by performing the Euclidean path integral, \((p-1)\)-form Maxwell theory on \(M_{d-1} = \S^1_\beta\times Y_{d-2}\), where \(\beta\) is the radius of \(\S^1\). Its action is simply
\begin{align}
	S[a] = \Half k\mu^\Delta\norm{f}^2 \coloneqq \Half k\mu^\Delta\int_{M} f\w\star f,
\end{align}
where \(k\) is a dimensionless constant, \(\mu\) is an energy scale and \(\Delta = d-1-2p\). \(f\in\Omega^p_\t{cl}(M)\) is a closed \(p\)-form on \(M\) and is identified with the curvature of a \((p-1)\)-form gauge field \(a\in \Omega^{p-1}(M)\). As such, it can be Hodge-decomposed as \(f = f_\t{h} + \dd{a}\), where \(f_\t{h}\in\harm^p(M)\) is a harmonic form, which we can (uniquely) choose to be orthogonal to \(\dd{a}\). \(f_\t{h}\) labels the instantons of the theory. Moreover, Dirac quantization condition, implies that \(f_\t{h}\) takes values in cohomology with coefficients in \(2\pi\Z\): \(f_\t{h}\in\H^p(M;2\pi\Z)\). The partition function of \((p-1)\)-form Maxwell theory on \(M\)  takes the form
\begin{equation}
	\parti^{(p-1)}_{\t{Maxwell}}[M] = \parti_\t{inst}[M]\parti_\t{osc}[M],
\end{equation}
where
\begin{align}
	\parti_\t{inst}[M] & \coloneqq \sum_{f_\t{h}\in\H^p(M;2\pi\Z)} \exp(-\Half k\mu^\Delta\norm{f_\t{h}}^2) \label{eq:Zinst-def-app}    \\
	\parti_\t{osc}[M]  & \coloneqq \int \frac{\DD{a}}{\vol(\cG_{p-1})} \exp(-\Half k\mu^\Delta\norm{\dd{a}}^2). \label{eq:Zosc-def-app}
\end{align}
In the above, \(\cG_{p-1}\) is the group of reducible gauge transformations, characteristic of higher-gauge theories, generated by shifts by closed \((p-1)\) forms, modulo their own gauge transformations.

In order to evaluate the above partition function, it will be helpful to introduce the topological basis of harmonic forms, \(\set{\tau_\sfi^{(k)}}_{\sfi=0}^{\b_k(M)}\) defined as in \Cref{eq:top-basis-def}. On \(M=\S^1_\beta\times Y\) this is
\begin{equation}\label{eq:top-basis-app}
	\set{\tau_\sfi^{(k)}}_{\sfi=0}^{\b_k(M)} = \set{\set{\overline{\tau}_\sfi^{(k)}}_{\sfi=0}^{\b_k(Y)},\set{\sigma\w\overline{\tau}_\sfi^{(k-1)}}_{\sfi=0}^{\b_k(Y)}}
\end{equation}
where \(\sigma \coloneqq \qty(2\pi \beta)^{-1} \vol_\beta\) is the unique normalised harmonic 1-form on the \(\S^1_\beta\). The Gram matrix for this basis:
\begin{align}
	\qty[\bbG_k^{M}]_{\sfi\sfj} \coloneqq \int_M \tau_\sfi^{(k)}\w \star \tau_\sfj^{(k)}
\end{align}
becomes, with the above decomposition
\begin{align}\label{eq:GM-decomp}
	\bbG_k^{M}=\mqty(\dmat{2\pi \beta \bbG_k^{Y}, (2\pi \beta)^{-1} \bbG_{k-1}^{Y}}),
\end{align}
with \(\bbG_k^Y\) being the analogous Gram matrix for \(Y\), defined over the basis \(\set{\overline{\tau}_\sfi^{(k)}}_{\sfi=0}^{\b_k(M)}\).

The instanton, \(f_\t{h}\) can be, therefore written as
\begin{align}
	f = 2\pi n^\sfi \overline{\tau}_\sfi^{(p)} + 2\pi m^\sfj \sigma\w\overline{\tau}_\sfj^{(p)},
\end{align}
with \(n^\sfi,m^\sfj\in\Z\). Combining \(n^\sfi\) and \(m^\sfj\) into vectors \(\vec{n}\) and \(\vec{m}\) of length \(\b_p(Y)\) and \(\b_{p-1}(Y)\) respectively, the instanton contribution, \Cref{eq:Zinst-def-app}, reads:
\begin{align}
	\parti_\t{inst}[M] = \sum_{\substack{\vec{n}\in\Z^{\b_p(Y)} \\ \vec{m}\in\Z^{\b_{p-1}(Y)}}} \exp(- \pi \mu^\Delta k\qty(2\pi\beta \vec{n}\cdot \bbG_p^Y \cdot\vec{n} + \frac{1}{2\pi \beta} \vec{m}\cdot \bbG_{p-1}^Y \cdot\vec{m})).
\end{align}
We can Poisson resum the sum over \(\vec{m}\) to obtain
\begin{equation}
	\parti_\t{inst}[M] = (2\pi\beta)^{\frac{1}{2}\b_{p-1}(Y)}\det(k \mu^\Delta \bbG_{p-1}^Y)^{-\frac{1}{2}} \Theta\qty[q;k\,\widetilde{\bbG}_p^Y]\Theta\qty[q;\inv{k}\,\widetilde{\bbG}_{d-p-1}^Y],
\end{equation}
where \(q\coloneqq \ex{- \beta \mu}\), the Siegel-type Theta functions are defined as in \Cref{eq:Theta}, the tilded versions of the Gram matrices are their dimensionless incarnations, having absorbed the relevant powers of \(\mu\), and we have also used the duality \(\bbG_k^Y = \inv{\qty[\bbG_{d-2-k}^Y]}\).

Coming to the oscillator contribution, this is given by \cite{Kelnhofer:2007jf,Szabo:2012hc,Donnelly:2016mlc}
\begin{align}
	\parti_\t{osc}[M] = \prod_{k=0}^{p-1} \det(\mu^\Delta k\, \bbG^M_k)^{\frac{1}{2}(-1)^{p-k-1}} \detp(\lapl_k)^{\frac{\kappa}{2}(p-k)(-1)^{p-k}},
\end{align}
where \(\lapl_k \coloneqq \dd\cdd+\cdd\dd\) is the Hodge Laplacian on \(k\)-forms on \(M\). and \(\detp\) excludes zero-modes. These are packaged in the first factor, the alternating product over \(\bbG_k^M\). In our case, \(M=\S^1_\beta\times Y\), we can vastly simplify this expression. Applying \Cref{eq:GM-decomp} in the zero-mode product we get
\begin{equation}
	\prod_{k=0}^{p-1} \det(\mu^\Delta k\, \bbG^M_k)^{\frac{1}{2}(-1)^{p-k-1}} = (2\pi\beta)^{-\frac{1}{2}\b_{p-1}(Y) + (-1)^{p-1}\sum_{n=0}^{p-2}(-1)^n\b_{n}(Y)} \det(k \mu^\Delta \bbG_{p-1}^Y)^{\frac{1}{2}}.
\end{equation}

Moving to the contribution of the Laplacian, note that its spectrum on \(M=\S^1_{\beta}\times Y\) is
\begin{align}
	\spec(\lapl_k) = \set{\qty(\frac{2\pi \sfn}{\beta})^2+\lambda_{\sfn_k},\ \sfn\in\Z,\ \sfn_k\in\sN_k}\cup \set{\qty(\frac{2\pi \sfn}{\beta})^2+\lambda_{\sfn_{k-1}},\ \sfn\in\Z,\ \sfn_{k-1}\in\sN_{k-1}}, \nonumber
\end{align}
where \(\sN_k\) is the index-set of the eigenvalues of the \(k\)-form Laplacian on \(Y\). From this it follows that the determinant of the Laplacian takes the form
\begin{align}
	\detp\lapl_k & = \underset{\scriptscriptstyle\qty(\frac{2\pi \sfn}{\beta})^2+\lambda_{\sfn_k}\neq 0}{\prod_{\sfn\in\Z}\prod_{\sfn_k\in\sN_k}} \qty(\qty(\frac{2\pi \sfn}{\beta})^2+\lambda_{\sfn_k})\ \gray{\times} \underset{\scriptscriptstyle\qty(\frac{2\pi \sfn}{\beta})^2+\lambda_{\sfn_{k-1}}\neq 0}{\prod_{\sfn\in\Z}\prod_{\sfn_{k-1}\in\sN_{k-1}}} \qty(\qty(\frac{2\pi \sfn}{\beta})^2+\lambda_{\sfn_{k-1}}) = \nn
	             & = \beta^{2\qty(\b_k(Y)+\b_{k-1}(Y))}\ \qty[\prod_{\sfn_k\in\sN_k^*}\sinh[2](\Half\beta\sqrt{\lambda_{\sfn_k}})]\ \prod_{\sfn_{k-1}\in\sN_{k-1}^*}\sinh[2](\Half\beta\sqrt{\lambda_{\sfn_{k-1}}}),
\end{align}
where we zeta-regularized the infinite products\footnote{Since we are interested in computing a path integral, we throw purely numerical coefficients, in particular various floating powers of 2, in this and the following expressions.}, using the spectral zeta-function, \(\upzeta_{Y}^{(k)}(s)\), for the \(k\)-form Laplacian on \(Y\), and \(\sN_k^*\) excludes the zero-modes. Lastly, note that \linebreak
\(\sN_k^* = \sN_k^\perp \oplus\sN_k^\parallel = \sN_k^\perp \oplus \sN_{k-1}^\perp,\)
meaning that the non-zero spectrum of the full Laplacian splits into a direct sum of the non-zero spectrum of the transversal Laplacian plus that of the longitudinal Laplacian. The latter is equal to the spectrum of the transversal Laplacian acting on \((k-1)\)-forms. With that, if we denote
\begin{align}
	S_k & \coloneqq \prod_{\sfn_k\in\sN_k^*}\sinh[2](\Half\beta\sqrt{\lambda_{\sfn_k}}) = \prod_{\sfn_k\in\sN_k^\perp}\sinh[2](\Half\beta\sqrt{\lambda_{\sfn_k}})\ \gray{\times} \prod_{\sfn_k\in\sN_{k-1}^\perp}\sinh[2](\Half\beta\sqrt{\lambda_{\sfn_{k-1}}}) \nn
	    & \eqqcolon S_k^\perp S_{k-1}^\perp, \nonumber
\end{align}
we have that
\begin{align}
	\prod_{k=0}^{p-1} \qty(\detp\lapl_k)^{\frac{1}{2}(p-k)(-1)^{p-k}} & = \beta^{(-1)^p \sum_{k=0}^{p-1}(-1)^k\b_k(Y)}\ \prod_{k=0}^{p-1} (S_k S_{k-1})^{\frac{\kappa}{2}(p-k)(-1)^{p-k}} = \nn
	                                                                  & = \beta^{(-1)^p \sum_{k=0}^{p-1}(-1)^k\b_k(Y)}\ \prod_{k=0}^{p-1} S_k^{\frac{\kappa}{2}(-1)^{p-k}} = \nn
	                                                                  & = \beta^{(-1)^p \sum_{k=0}^{p-1}(-1)^k\b_k(Y)}\ \qty(S_{p-1}^\perp)^{-\frac{\kappa}{2}} = \nn
	                                                                  & = \beta^{(-1)^p \sum_{k=0}^{p-1}(-1)^k\b_k(Y)}\ \qty(\upeta_{Y}^{(p-1)}[q])^{-2},
\end{align}
where we used \Cref{eq:eta-def2}, to express the result in terms of the \((p-1)\)-form \(\eta\) function associated with \(Y\). Putting everything together, the oscillator contribution to the partition function is
\begin{align}\label{eq:parti-osc-app}
	\parti_\t{osc}[M] = (2\pi\beta)^{\frac{1}{2}\b_{p-1}(Y)} \det(k \mu^\Delta \bbG_{p-1}^Y)^{\frac{1}{2}}\ \qty(\upeta_{Y}^{(p-1)}[q])^{-2},
\end{align}
Combining with the instanton contribution, all the factors of \(\beta\) and \(k\) exactly cancel and we get:
\begin{empheq}[box=\obox]{equation}
	\parti^{(p-1)}_{\t{Maxwell}}\qty[\S^1_\beta\times Y] = \frac{\Theta\qty[q;k\,\widetilde{\bbG}_p^Y]}{\upeta_{Y}^{(p-1)}[q]}\ \frac{\Theta\qty[q;\inv{k}\,\widetilde{\bbG}_{p-1}^Y]}{\upeta_{Y}^{(d-p-2)}[q]},
\end{empheq}
where we also used the fact that the eta function associated with $(p-1)$-forms is equal to that for $(d-p-2)$-forms, in order to write the result in a symmetric fashion.

\subsection{The partition function of chiral mixed Maxwell theory}

Now that we have all the details of the vanilla, higher-form gauge theory under our belt, we will quantize the chiral mixed Maxwell theory on \(M = \S^1_\beta\times Y\). This is a theory with action
\begin{align}
    S[\cA] = \frac{k \mu^\Delta}{2}\norm{\cA}^2 = \frac{k \mu^{-\Delta}}{2} \norm{\cB}^2,
\end{align}
where \(\cA\) and \(\cB\) are closed \(p\)- and \((d-p-1)\)-forms, respectively, satisfying the generalized chiral condition \(\cB + \mu^\Delta (-1)^{(d-p-1)(p+1)} \star \cA \demeqq 0\). In the main text we argued that this leads to fractionally quantized fluxes along cycles involving the thermal circle. Let us elaborate on that. Dirac quantization imposes that magnetic and electric fluxes are quantized as:
\begin{align}
    \Phi^\cA_\t{mag}[\eta] \coloneqq\frac{1}{2\pi}\int_\eta \cA \in \Z \qquad \Phi^\cA_\t{elec}[\gamma] \coloneqq \frac{k}{2\pi}\int_\gamma \star \cA \in \Z \\
    \Phi^\cB_\t{mag}[\gamma] \coloneqq\frac{1}{2\pi}\int_\gamma \cB \in \Z \qquad \Phi^\cB_\t{elec}[\eta] \coloneqq\frac{k}{2\pi}\int_\eta \star \cB \in \Z,
\end{align}  
for all cycles \(\eta\in\H_p(Y)\) and \(\gamma\in\H_{d-p-1}(Y)\). Note that the electric fluxes are defined including a factor of the coupling constant, here \(k\) \cite{Verlinde:1995mz}. The generalized chiral condition then requires that magnetic fluxes of \(\cA\) along cycles of the form \(\S^1_\beta\times \widetilde{\eta}\), for \(\widetilde{\eta}\in\H_{p-1}(Y)\) are quantized in units of \(\frac{1}{k}\). Consequently, decomposing \(\cA\), as in the Maxwell story, as \(\cA = \cA_\t{h} + \dd{a}\), where \(\cA_\t{h}\) is harmonic and then further decomposing \(\cA_\t{h}\) in terms of the basis \Cref{eq:top-basis-app}, we get
\begin{equation}
    \cA_\t{h} = 2\pi n^\sfi \overline{\tau}_\sfi^{(p)} + \frac{2\pi}{k} m^\sfj \sigma\w\overline{\tau}_\sfj^{(p)}.
\end{equation}
From here on, the story is completely analogous to the vanilla Maxwell case. The instanton contribution is
\begin{align}
    \parti_\t{inst}\qty[M] = (2\pi\beta)^{\frac{1}{2}\b_{p-1}(Y)}\det(k \mu^\Delta \bbG_{p-1}^Y)^{-\frac{1}{2}} \Theta\qty[q;k\,\widetilde{\bbG}_p^Y]\ \Theta\qty[q; k\,\widetilde{\bbG}_{d-p-1}^Y].
\end{align}

The oscillators are completely unaffected by this story. It is cleaner to write their contribution in terms of the magnetic photon, \(\widetilde{a}\), such that \(\dd{\widetilde{a}}= k\mu^{-\Delta}\star\dd{a}\). It is, of course, not necessary to do that; this way we will just avoid the necessity of adding a counterterm. Doing so results in an oscillator contribution as 
\begin{align}
    \parti_\t{osc}[M] = (2\pi\beta)^{-\frac{1}{2}\b_{p-1}(Y)} \det(k \mu^\Delta \bbG_{p-1}^Y)^{\frac{1}{2}}\ \qty(\upeta_{Y}^{(p-1)}[q])^{-2}.
\end{align}
In total, the partition function reads:
\begin{empheq}[box=\obox]{equation}
    \parti^{(p-1,d-p-2)}_{\t{chiral}}\qty[\S^1_\beta\times Y] = \frac{\Theta\qty[q;k\,\widetilde{\bbG}_p^Y]}{\upeta_{Y}^{(p-1)}[q]}\ \frac{\Theta\qty[q;k\,\widetilde{\bbG}_{d-p-1}^Y]}{\upeta_{Y}^{(d-p-2)}[q]}.
\end{empheq}

\section{Laplacians, zeta functions, and heat kernels}\label{app:zeta}

In this appendix, we discuss some spectral properties of the Hodge Laplacian. 

Let \((Y,g)\) be a closed, compact, \((d-2)\)-dimensional Riemannian manifold and let \linebreak \(\lapl_\ell \coloneqq \cdd\dd+\dd\cdd\) be the Hodge Laplacian acting on \(\Omega^\ell(Y)\). Denoting by \(\spec'\) the non-zero-mode spectrum, it holds that
\begin{align}
	\spec'\qty(\lapl_\ell,X) & = \spec'\qty(\lapl_\ell\Big|_{\ker\cdd},Y)\oplus\spec'\qty(\lapl_\ell\Big|_{\ker\dd},Y) = \nn
	                         & = \spec'\qty(\lapl_\ell\Big|_{\ker\cdd},Y)\oplus\spec'\qty(\lapl_{\ell-1}\Big|_{\ker\cdd},Y).
\end{align}
Therefore, we will focus on the spectral properties of the operator \(\lapl_\ell\Big|_{\ker\cdd}\) which we will call \(\tlapl_\ell\). Consider the eigenvalue equation for \(\tlapl_\ell\):
\begin{align}
	\tlapl_\ell \varphi_\sfn = \lambda_\sfn \varphi_\sfn, \qquad \sfn\in\sN_\ell,
\end{align}
with \(\sN_\ell\) being a countable set. We can define the spectral zeta function of \(\tlapl_\ell\):
\begin{align}
	\sz{W}{\ell}(s) \coloneqq \sum_{\sfn\in\sN_\ell^*} \lambda_\sfn^{-s}, \qquad s\in\C,\ \t{with}\ \Re s >\frac{d-2}{2}, \label{eq:MP-zeta}
\end{align}
where \(\sN_\ell^*\coloneqq\set{\sfn\in\sN_\ell\suchthat \lambda_\sfn\neq 0}\). For \(\ell=0\) this reduces to the well-known Minakshisundaram--Pleijel zeta function \cite{minakshisundaram_pleijel_1949}.

It will be useful to introduce the heat kernel:
\begin{align}
	\cK_Y^{(\ell)}(t) \coloneqq \tr\qty(\ex{-t \tlapl_p}) = \sum_{\sfn\in\sN_\ell} \ex{-t \lambda_\sfn}.
\end{align}
The heat kernel admits a small-\(t\) expansion, as
\begin{align}
	\cK_Y^{(\ell)}(t) = \frac{1}{\qty(4\pi t)^{\frac{d-2}{2}}}\sum_{k=0}^\infty \cI_k^{(\ell)} t^k, \label{eq:K-expansion}
\end{align}
where \(\cI_k^{(\ell)}\) are given by integrals of geometric data of \(W\). We can now invoke a Mellin transform in order to write
\begin{align}
	\lambda_\sfn^{-s} = \frac{1}{\Gamma(s)}\int_{0}^\infty \dd{t} \ex{-t \lambda_\sfn} t^{s-1},
\end{align}
and hence
\begin{align}
	\sz{W}{\ell}(s) = \frac{1}{\Gamma(s)}\int_0^\infty \dd{t} t^{s-1} \qty(\cK_Y^{(\ell)}(t)-\dim\ker\tlapl_\ell) \eqqcolon \frac{1}{\Gamma(s)}\int_0^\infty \dd{t} t^{s-1}\; \widetilde{\cK}_Y^{(\ell)}(t).
\end{align}
where \(\dim\ker\tlapl_\ell\) counts the number of of zero-modes.

We will now use the following trick. We will split the above integral as follows.
\begin{align}
	\sz{W}{\ell}(s) = \frac{1}{\Gamma(s)}\int_0^\infty \dd{t} \Theta(t-1)\,t^{s-1}\; \widetilde{\cK}_Y^{(\ell)}(t) + \frac{1}{\Gamma(s)}\int_0^\infty \dd{t}\qty(1-\Theta(t-1))\,t^{s-1}\; \widetilde{\cK}_Y^{(\ell)}(t) \label{eq:z-splitting}
\end{align}
where \(\Theta(t)\) is the Heaviside Theta function.\footnote{For continuity purposes, one could use a smoothened version of the Heaviside Theta function.} The merit of this is that the second integral in \Cref{eq:z-splitting} is now manifestly analytic in \(s\), while the first integral is explicitly computable. Plugging in the asymptotic expansion, \Cref{eq:K-expansion}, of \(\cK_Y^{(\ell)}\), into the first integral, one gets
\begin{align}
	\sz{Y}{\ell}(s) = \sum_{k=0}^\infty \frac{\cI_{k}^{(\ell)}}{\qty(4\pi)^{\frac{d-2}{2}}\Gamma(s)}\ \frac{1}{s-\frac{d-2}{2}+k} + \frac{\dim\ker\tlapl_\ell}{s \Gamma(s)} - \frac{H(s)}{\Gamma(s)}, \label{eq:zeta-poles}
\end{align}
where \(H(s)\) is analytic in \(s\).

From this we get the pole structure of the spectral zeta function. It has simple poles at \(s=\frac{d-2}{2}-k\), with \(k\in\Z_{\geq 0}\), with residue
\begin{align}
	\Res\sz{Y}{\ell}\qty(\frac{d-2}{2}-k) = \frac{\cI^{(\ell)}_{k}}{\qty(4\pi)^{\frac{d-2}{2}}\Gamma\qty(\frac{d-2}{2}-k)}.
\end{align}
When \(d\) is even, the spectral zeta-function has poles only at \(s>0\), i.e. \(k\in\set{0,1,\cdots \frac{d-2}{2}-1}\), because of the Gamma function in the denominator.

Moreover, notice that at \(s=0\), we have
\begin{align}
	\sz{Y}{\ell}(0) =
	\begin{cases}
		\cI^{(\ell)}_{\frac{d-2}{2}} - \dim\ker\tlapl_\ell,      & \t{when \(d\) is even} \\
		\phantom{\cI^{(p)}_{\frac{d-2}{2}}~}- \dim\ker\tlapl_\ell, & \t{when \(d\) is odd}.
	\end{cases}
\end{align}
This follows immediately from \(\Gamma(s)\sim\frac{1}{s}\) as \(s\to 0\), since the analytic piece \(H(s)/\Gamma(s)\) vanishes as \(s\to 0\); the topological piece, \(-\dim\ker\tlapl_\ell\), is universal, and the first term only contributes when \(\ell\) is even. Finally we have that \(\dim\ker\tlapl_\ell = \dim\ker\lapl_\ell = \b_\ell(Y)\), so
\begin{align}
	\sz{Y}{\ell}(0) =
	\begin{cases}
		\cI^{(\ell)}_{\frac{d-2}{2}} - \b_\ell(Y),       & \t{when \(d\) is even} \\
		\phantom{\cI^{(p)}_{\frac{d-2}{2}}~} - \b_\ell(Y), & \t{when \(d\) is odd}.
	\end{cases}
\end{align}

\printbibliography
\end{document}